# Model predictive control of resistive wall mode for ITER

Samo Gerkšič[a], Boštjan Pregelj[a], and Marco Ariola[b]

[a]*Jožef Stefan Institute, Jamova 39, SI-1000 Ljubljana, Slovenia*
[b]*Consorzio CREATE / Università degli Studi di Napoli Parthenope, Naples, Italy*

Abstract: Active feedback stabilization of the dominant resistive wall mode (RWM) for an ITER H-mode scenario at high plasma pressure using infinite-horizon model predictive control (MPC) is presented. The MPC approach is closely-related to linear-quadratic-Gaussian (LQG) control, improving the performance in the vicinity of constraints. The control-oriented model for MPC is obtained with model reduction from a high-dimensional model produced by CarMa code. Due to the limited time for on-line optimization, a suitable MPC formulation considering only input (coil voltage) constraints is chosen, and the primal fast gradient method is used for solving the associated quadratic programming problem. The performance is evaluated in simulation in comparison to LQG control. Sensitivity to noise, robustness to changes of unstable RWM dynamics, and size of the domain of attraction of the initial conditions of the unstable modes are examined.

Keywords: predictive control, plasma magnetic control, quadratic programming, fast gradient method.

## 1. Introduction

The resistive wall mode (RWM) is a helical deformation of predominantly axi-symmetric plasma in toroidal plasma devices, a magneto-hydrodynamic (MHD) instability which arises due to the kink of pressurized plasma at characteristic rational surfaces (*n*, *m*), where *n* and *m* are the toroidal and the poloidal mode number, respectively, and is moderated by the conductive reactor wall [1, 2, 3]. The external kink instability is inherent to plasma MHD equilibria at normalized plasma pressures $\beta_N$ above the so-called "no-wall" limit. Without a conductive wall, the kink instability would appear with a very fast Alfvénic MHD growth time $\tau_A$ in the μs range, beyond reach of magnetic feedback control. In the presence of a nearby infinitely-conductive wall, the currents induced in the wall counteract the kink, and the stable $\beta_N$ range is extended to the higher "ideal-wall" limit. With a wall of finite resistivity, the currents induced in the wall turn the fast kink into a RWM with the wall growth time $\tau_W$ in the ms range. But at higher plasma pressures above the no-wall $\beta_N$, the RWM dynamics are still unstable. Passive stabilization may be possible by plasma rotation (via neutral-beam heating/ECCD or rotating magnetic field perturbation) or kinetic effects [4, 5, 6, 7], however these may not be sufficient in reactor conditions expected to be suitable for exploiting nuclear fusion for commercial-scale energy production in long-pulse high-$\beta_N$ H-mode scenarios at the fusion energy gain factor *Q* exceeding 5. Hence, active RWM feedback control via corrective magnetic coils, which augment the effect of the conductive wall and attempt to provide balance to the RWM magnetic disturbance, is planned for advanced ITER experiments, and is one of the important open scientific questions on the path towards fusion energy [5, 8, 9, 10].

Extensive research of RWMs was carried out in reversed-field pinch (RFP) and tokamak experiments in RFX-mod [11, 12, 13, 14, 15], EXTRAP T2R [16, 17, 18, 19, 20], HBT-EP [21], DIII-D [4, 22, 23, 8, 24, 15, 25, 26, 27], NSTX [28, 29, 30], and JT-60 [6]. Modelling codes, such as VALEN [31, 26, 27], CarMa [32, 33, 34, 35, 36], CarMa0NL [37, 38] are generally based on coupling MHD equations for the plasma with the equations describing the currents induced in the conductive wall. RWM control research for ITER [39, 40, 10] currently relies on modelling. Experimentation at high $\beta_N$ is not planned during the initial stages of ITER operation, because disruptions may cause substantial damage to the reactor wall elements.

The majority of past RWM control experiments focused on current-driven RWMs, mainly because they may be repeatably triggered with fast current ramp-up [15], or on disturbances induced with a different set of coils [26, 27]. Pressure-driven RWMs [29] are more relevant to advanced tokamak scenarios, however they are less reproducible and often accompanied with other types of MHD instabilities, such as edge localized modes (ELM) and neoclassical tearing mode (NTM) [1, 3, 6, 8, 23, 21, 41].

Magnetic active feedback RWM control is closely related with error field compensation (EFC) [4]. EFC is intended for the correction of both the static (vacuum) error field due to the imperfections of the main magnetic coils and the varying resonant perturbation of the plasma to the remaining error field, which cannot be fully compensated due to geometry limitations [38, 42]. Both EFC and active RWM feedback act on correction coils positioned perpendicularly along the reactor wall, either internally or externally to the vacuum vessel. They may be implemented using separate coil sets [22, 15] or the same coil set [4, 24, 8, 29, 17, 20, 18, 19], in some cases even within the same controller [17, 20, 18, 19]. The main distinction is in dynamics – EFC acts on the static field and the slow time-variation of the plasma response

___________________________________________________________________

*Corresponding author's email: Samo.Gerksic@ijs.si*

to it, while active RWM feedback takes care of dynamics in the range of $\tau_W$ and must be able to stabilize the open-loop-unstable RWM. Internal coils are generally preferred for active RWM control because of faster response. In ITER, EFC is planned to be actuated using three (top, side, bottom) sets of 6 external superconducting correction coils with 9 independent power supplies [43, 44, 45], while active RWM feedback may be implemented as a secondary function using three sets of 9 internal ELM coils [46, 47] (which are primarily intended for suppression other MHD instabilities via rotating magnetic field perturbation [48, 49]). EFC is outside the scope of this work, and it is assumed that a separate EFC system will be used, but it may be interesting to explore the interaction of EFC and active RWM for simultaneous multiple mission sharing [39] in more detail in future.

The initial active RWM feedback approaches were mostly based on multiple local single-input single-output (SISO) proportional (P) or proportional-derivative (PD) control of actuator coils based on adjacent sensor measurements of the radial and/or poloidal magnetic field [4, 22, 16, 8, 14, 25, 5], possibly also including a toroidal phase shift [28, 8, 15]. Subsequently, improvements of the control performance were shown with advanced model-based controllers, such as Kalman-filter-based estimation of the kink mode with P/PD feedback [23, 21], linear quadratic Gaussian (LQG) control [50, 51, 31, 29, 17, 46, 47, 26], robust $H_\infty$ control [52, 53], and model predictive control (MPC) [18, 19]. With these approaches, the controller is designed based on a control-oriented model. This model may be obtained either with model identification procedures from experimental data, or with first-principles modelling and model simplification/reduction procedures; sometimes the two approaches are combined.

LQG control is based on the linear quadratic (LQ) optimal controller, with a quadratic cost function penalizing the system state and the control signal over the infinite future horizon, typically a using linear state-space model [50, 51, 31]. The minimization of the cost function is solved using the Riccati equation, and in the most commonly used time-invariant form, a matrix-vector multiplication of the constant controller gain matrix with the current system state vector is required for the computation of the controller output in a particular time instant. The system state vector is estimated from system output measurements with Gaussian noise assumption using an optimal state estimator known as the Kalman filter (KF) which is dual to the LQ control problem [50, 51, 31]. The initial LQG optimal control designs for RWM control were based on low-dimensional SISO models [50, 51]. Later approaches mostly use use multiple-input multiple-output (MIMO) models where interactions are considered more systematically and RWMs are treated as global states. Katsuro-Hopkins et al. [31] presented LQG control for ITER based on the VALEN model, also applied to NSTX [29]. Villone and Ariola [46, 47] implemented LQG control for ITER based on the CarMa model, with emphasis on RWM interaction with vertical stabilization (VS, control of the unstable axi-symmetrical $n = 0$ mode) [46]. Clement et al. [26, 27] presented a fast LQG implementation and experimental evaluation on the DIII-D based on the VALEN model, showing that LQG feedback using external control coils may be as effective as proportional feedback using internal control coils. The advantage of LQG over PD control was also shown by Dalessio et al. [52, 53] where control design was further upgraded with the $H_\infty$ control approach maximizing robustness to model uncertainty expressed as a single structured uncertainty parameter addressing the variable RWM growth rate. Setiadi et al. [20] presented a LQG controller for EXTRAP T2R with a gray-box modelling approach combining a finite-element model for the conductive shell and black-box modelling (model identification) for the plasma. The volume of experimental results of LQG with high-$\beta_N$ H-mode tokamak plasmas is relatively limited in comparison to the conventional PD-control-based approaches, but it was demonstrated that LQG may facilitate longer pulse duration in H-mode high $\beta_N$ conditions in NSTX [30] and DIII-D [26, 27].

Villone and Pironti [54] and Ariola and Pironti [47] point out the issue of best achievable performance under constraints using LQG for ITER VS and RWM control, respectively. However, the performance of such control in the presence of constraints is not optimal, because the LQ optimal controller does not take the presence of constraints into account. The constraints are actively considered with typical implementations of the closely related MPC approach [55, 56] (in a particular infinite-horizon setup, the two are blended in constrained LQ control [57]). The implementation of MPC requires on-line optimization, typically in the form of quadratic programming (QP) [55, 56, 58]. Therefore, MPC for MIMO systems used to be restricted to processes with relatively slow dynamics, but recently a considerable progress in fast online QP solvers was made [59, 60, 61, 62, 63, 64]. This enabled several applications of MPC in magnetic control for tokamaks and RFPs. Maaljars et al. [65, 66] presented MPC control of the plasma pressure and safety factor profile for ITER and TCV using the RAPTOR code; Wehner et al. [67] uses MPC for control of plasma safety factor profile for DIII-D. Gerkšič et al. [68, 69] implemented MPC plasma current and shape control for ITER using a dual fast gradient method (FGM) QP solver [63]. Gerkšič et al. [70] used explicit MPC for VS of the $n = 0$ mode for ITER, which does not require on-line optimization because a parametric solution is computed in advance, but this is suitable only for low-dimensional control problems.

Closely related work by Setiadi et al [18] on fast MPC control of RWMs for EXTRAP T2R using black-box modelling shows that a primal FGM QP solver is computationally feasible even for devices with much faster dynamic than ITER, albeit with a very short predictive horizon, and it does not explore the influence of constraints handling and the relation to the LQG control. In [19] the approach is extended for EFC with two input disturbance estimation approaches that, unlike more conventional MPC disturbance estimation approaches [56], allow integral action without augmenting the control model.

In this work we present active feedback RWM stabilization for ITER using infinite-horizon MPC with a reduced-order state-space linear model derived using CarMa code. The approach follows the LQG approach of Ariola and Pironti [47]. The controller is designed for the MHD plasma equilibrium of an ITER high-$β_N$ H-mode scenario that approaches the conditions deemed appropriate for fusion energy production, where $β_N$ slightly exceeds the no-wall limit, so that one unstable dominant $n = 1$ RWM is expected. We show that the solution to the on-line optimization problem of the MPC controller is computationally feasible within the very short available time, using a standard computer. We choose a suitable MPC formulation with input constrains only [67, 66, 18, 62], and the resulting QP is solved using the primal FGM method [63, 72]. We demonstrate the efficiency of handling of coil voltage constraints by MPC compared to LQG control with and without estimator wind-up protection and its impact on best achievable performance under constraints. We demonstrate the robustness of the control system to variation of RWM parameters; in particular, despite the controller being designed for a specific MHD equilibrium, the control responses are well-behaved for systems with actual growth rates lower than the nominal one, including the stable RWM case.

In Section 2 we describe the ITER RWM control problem setup. Section 3 describes the design of the proposed infinite-horizon MPC controller with the KF, a performance comparison with a closely related LQG controller, and explore its sensitivity to noise, its robustness to changes of RWM parameters, and its best achievable performance with respect to the stabilizable region of initial values of two unstable modes. The description how the MPC control problem is efficiently solved with online optimisation using the primal FGM method is provided in Section 4.

## 2. RWM control for ITER

We are considering the problem of active feedback stabilization of the dominant $n = 1$ RWM in ITER, where active stabilization of an open-loop-unstable RWM is needed. The control problem setup follows the description of Ariola and Pironti [47]. The in-vessel copper coils are shown in Figure 1.

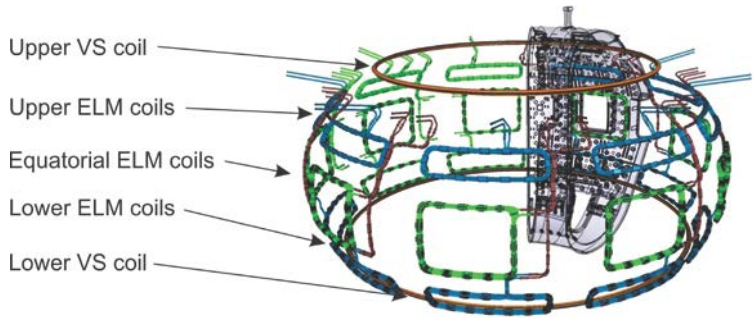

Figure 1. ITER In-vessel coils: axi-symmetric VS coils and non-axi-symmetric ELM coils). Credit © ITER Organization, http://www.iter.org/

In this work, 27 non-axi-symmetric ELM coils are used for RWM control. The ELM coils are arranged in three sectors vertically (upper, equatorial, lower) and nine sectors horizontally along the toroidal chamber; the horizontal sectors are equally-spaced and located at the toroidal angles $φ_i = 40° · (i – 1)$, with $i = 1, ..., 9$. In order to facilitate handling of actuator constraints, the ELM coils with their corresponding power supplies are treated as 27 independent actuators, without applying reduction of actuator vector dimension as in [47]. The control signal vector applied to the inputs of the coil power supplies (PS) is denoted as $\mathbf{u} \in \Re^{27}$, and the vector of output PS voltages fed to the ELM coils is $\mathbf{u}_{ELM} \in \Re^{27}$. For control design, each of the power supplies is modelled as a saturation block, with the limits $V_{max} = 144$ V and $V_{min} = – V_{max}$, and a first-order transfer function with time delay[1]

$$G_{PS}(s) = \frac{1}{7.5 \cdot 10^{-3} s + 1} e^{-2.5 \cdot 10^{-3} s} \quad (1)$$

The absolute value of the ELM coil currents should not exceed $1.5 \cdot 10^4$ A. The design limits of voltages and currents of the ELM coils are much higher, however RWM suppression is their secondary function.

The measurement set used for the RWM stabilization comprises 6 vertical magnetic field $B_v$ sensors located at the same poloidal positions ($r = 8.928$ m, $z = 0.550$ m) toroidally spaced at approximately 60° angles apart, $φ_i \in \{39°, 101°, 159°, 221°, 279°, 341°\}$ comprises the measurement output signal $\mathbf{y}_m \in \Re^6$. The control scheme is intended for suppression of oscillating disturbances that may be evaluated as

$$y_i = y_A \cos φ_i + y_B \sin φ_i , \quad i \in \{1, 2, ..., 6\} \quad (2)$$

---

[1] The values of the parameters are assumptions, as the power supply design for the ELM correction coils is not finalized yet [10, 71].

so that a reduced-dimensional measurement vector $\mathbf{y} \in \Re^2$ containing the cosine and sine components of the disturbance [47] is determined with the minimum-mean-square formula (where $\dagger$ denotes the Moore-Penrose inverse)

$$\mathbf{y} = \begin{bmatrix} y_A \\ y_B \end{bmatrix} = \mathbf{T}_{\text{out}}\mathbf{y}_m, \quad \mathbf{T}_{\text{out}} = \begin{bmatrix} \cos\varphi_1 & \sin\varphi_1 \\ \cos\varphi_2 & \sin\varphi_2 \\ \vdots & \vdots \\ \cos\varphi_6 & \sin\varphi_6 \end{bmatrix}^{\dagger} \quad (3)$$

Our MPC control design is based on the CarMa modelling approach [32, 33, 34 35]. It is assumed that the relevant RWM dynamics are captured by using three-dimensional (3D) axi-symmetric linearised MHD equations for the plasma, solved by the MARS-F code, and equations describing the currents induced in the 3D volumetric conductors of the wall surrounding the plasma, discretized to a mesh of hexahedral elements, solved using the 3D eddy current integral formulation of the CARIDDI code. The plasma and wall equations are combined by evaluating their electromagnetic contributions on a chosen *coupling surface* in between the plasma and the surrounding conductive structures.

By means of the CarMa modelling code, the following linear state-space model describing perturbations from the equilibrium configuration for the plasma current $I_p$ = 9 MA and the normalized beta $\beta_N$ = 2.85 is obtained (all values are displacements from the equilibrium although the difference operator $\Delta$ is omitted)

$$\dot{\mathbf{x}}_{CM} = \mathbf{A}_{CM}\mathbf{x}_{CM} + \mathbf{B}_{CM}\mathbf{u}_{ELM} \quad (4)$$

$$\mathbf{y}_m = \mathbf{C}_{CM,m}\mathbf{x}_{CM} \quad (5)$$

where the elements of the state vector $\mathbf{x}_{CM} \in \Re^{6294}$ coincides with the set of displacements of the 3D currents in the structure, and $\mathbf{A}_{CM}$, $\mathbf{B}_{CM}$, $\mathbf{C}_{CM,m}$ are the dynamic, input and output matrix of the RWM dynamics state-space model, respectively. The high model order results from the 3D mesh of the structure that considers: the vessel without the ports, the outer triangular support, the copper cladding, and the vertical stabilization in-vessel coils; see [34] for further considerations regarding modelling of the conductive structure. The model also includes an auxiliary output vector of ELM coil currents $\mathbf{I}_{ELM} \in \Re^{27}$ via the auxiliary output matrix $\mathbf{C}_{CM,ELM}$

$$\mathbf{I}_{ELM} = \mathbf{C}_{CM,ELM}\mathbf{x}_{CM} \quad (6)$$

Each of the eigenvectors of the dynamic matrix $\mathbf{A}_{CM}$ corresponds to a specific current pattern inside the 3D structure, taking into account the presence of the plasma [47]. $\mathbf{A}_{CM}$ has two similar eigenvalues corresponding to the unstable $n$ = 1 RWM mode[2], with the growth rate $\gamma$ (about 19 s$^{-1}$) and frequency $\omega$ (about 0.26 rad/s). With a state-space transformation to the *modal form* $\boldsymbol{\xi} = \mathbf{T}_M \mathbf{x}_{CM}$, where $\boldsymbol{\xi} \in \Re^{6294}$ is the transformed state vector, the state-transition equation (4) can be reorganized with the transformed dynamic matrix $\mathbf{A}_{CMT}$ decomposed to the unstable and stable parts as

$$\begin{bmatrix} \dot{\xi}_1 \\ \dot{\xi}_2 \\ \dot{\boldsymbol{\xi}}_s \end{bmatrix} = \begin{bmatrix} \gamma & \omega & \mathbf{0} \\ -\omega & \gamma & \mathbf{0} \\ \mathbf{0} & \mathbf{0} & \mathbf{A}_{CMT,s} \end{bmatrix} \begin{bmatrix} \xi_1 \\ \xi_2 \\ \boldsymbol{\xi}_s \end{bmatrix} + \begin{bmatrix} \mathbf{B}_{CM,u} \\ \mathbf{B}_{CM,s} \end{bmatrix} \mathbf{u}_{ELM} \quad (7)$$

where the indices $u$ and $s$ denote the unstable and stable dynamics of the model, respectively, $\mathbf{0}$ denotes a vector or matrix of appropriate dimensions with all zero elements, and $\xi_1$ and $\xi_2$ are the cosine and sine components of the unstable $n$ = 1 RWM mode. The model order is high because of the fine mesh used for the discretisation of the 3D volumetric conductors; however, it is neither feasible nor required to control the stable modes individually, because the spatial distribution is not relevant for control.

## 3. Model Predictive Control

The introduction of MPC to RWM control in this work is aimed at improving the control performance in the areas close to the actuator coil voltage constraints. Using a fast QP solver such as the primal FGM method, this is deemed computationally feasible at a sub-millisecond sampling time required for ITER, and with more efficient FPGA computational implementation also at shorter sampling times as needed for experimental validation on dynamically faster smaller-scale tokamaks.

For the implementation of a model-based controller, a control-oriented model of the controlled system comprising the PS dynamics (2) with Padé approximation of time-delay and the RWM dynamics (4-5) in a suitable form is needed. For the considered MPC approach, a discrete-time state-space model is required

$$\mathbf{x}(k+1) = \mathbf{A}\mathbf{x}(k) + \mathbf{B}\mathbf{u}(k) \quad (8)$$

---

[2] Generally, the CarMa modelling approach may consider several RWMs with different mode numbers, and also MPC/LQG control may be extended to several unstable modes.

$$\mathbf{y}(k) = \mathbf{C}\mathbf{x}(k) \qquad (9)$$

where $k$ is the discrete time index, $\mathbf{A}$, $\mathbf{B}$, and $\mathbf{C}$ are the system matrices obtained via zero-order-hold discretisation with the sampling time $T_s = 0.75$ ms (as in [47]), and $\mathbf{x} \in \Re^{N_x}$ is the state vector, with the model order $N_x$ reduced as much as possible while still adequately covering the relevant frequency range of dynamics. A comparison of Bode diagrams of the original continuous-time model (black) and the discretised reduced-order model (8-9) (cyan) is shown in Figure 2. For numerical reason, the improved Davison method [73] is used firstly to reduce the order of the RWM dynamics from several thousands to several hundreds. In the displayed range, the first reduction matches the original model. Then, the balanced truncation technique [74] is used to reduce the order of the final model to 50. The Bode diagram shows that the reduced-order model matches the original model well in the lower and medium frequency range up to around 100 rad/s. With higher-order models, better matching at higher frequencies is achieved, however control simulations (with the LQG and MPC controllers as designed and tuned in the following) do not show a significant performance improvement. Similarly, control simulations indicate that the performance could be slightly improved with a shorter sampling time, and that longer sampling times lead to a considerable performance deterioration.

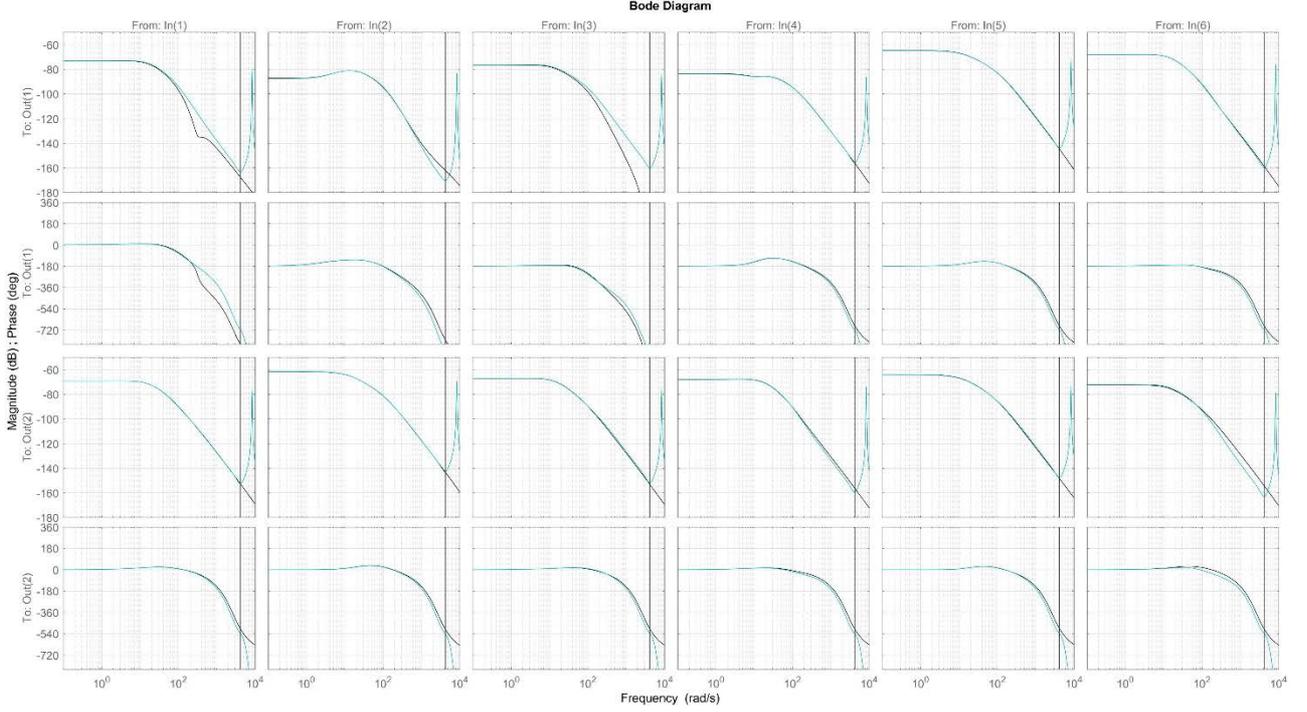

Figure 2. Bode diagram of the controlled system model: original (black) and the reduced-order discretised model (cyan). The responses from the first 6 (out of 27) elements of $\mathbf{u}$ to both outputs $\mathbf{y}$ are shown. The black vertical line marks the Nyquist frequency of the discrete-time reduced-order model.

An infinite-horizon MPC controller is proposed. It is based on a cost function consisting of two parts: a finite-horizon MPC cost for the first part of the future horizon from sample $k$ up to sample $(k + N - 1)$, and a terminal infinite-horizon LQ cost for samples from $N$ onwards, where $N$ denotes the length of the finite horizon. Constraints are actively considered only in the first part of the horizon. The cost function is

$$J(k) = \tfrac{1}{2}\sum_{i=0}^{N-1}\left(\mathbf{x}_{k+i}^T \mathbf{Q}_C \mathbf{x}_{k+i} + \mathbf{u}_{k+i}^T \mathbf{R}_C \mathbf{u}_{k+i}\right) + \tfrac{1}{2}\mathbf{x}_{k+N}^T \mathbf{P} \mathbf{x}_{k+N} \qquad (10)$$

where the notation using lower discrete-time indices denotes signal prediction based on the current state $\mathbf{x}(k)$; $\mathbf{Q}_C$ and $\mathbf{R}_C$ are state and control cost matrices, respectively; the term $\mathbf{x}_{k+N}^T \mathbf{P} \mathbf{x}_{k+N}$ represents the terminal LQ cost, and $\mathbf{P}$ is the steady-state solution of the discrete-time algebraic Riccati equation [74]

$$\mathbf{P} = \mathbf{A}^T \mathbf{P} \mathbf{A} - \mathbf{A}^T \mathbf{P} \mathbf{B}(\mathbf{R}_C + \mathbf{B}^T \mathbf{P} \mathbf{B})^{-1} \mathbf{B}^T \mathbf{P} \mathbf{A} + \mathbf{Q}_C \qquad (11)$$

computed using the Matlab function `lqr`.

The control signal is computed by solving the MPC optimization problem

$$\min_{\mathbf{u}_k,\ldots,\mathbf{u}_{k+N-1}} J(k)$$

$$\text{subject to } \mathbf{x}_{k+j+1} = \mathbf{A}\mathbf{x}_{k+j} + \mathbf{B}\mathbf{u}_{k+j}, \quad j = 0, 1, \ldots, N-1$$

$$\mathbf{u}_{\min} \le \mathbf{u}_{k+j} \le \mathbf{u}_{\max}, \quad j = 0, 1, \ldots, N-1$$

$$\mathbf{x}_k = \mathbf{x}(k) \qquad (12)$$

The system state $\mathbf{x}(k)$ is not measurable, so in its place its current estimate $\mathbf{x}(k|k)$ is used, which is obtained using the steady-state Kalman filter (KF) [74]

$$\mathbf{x}(k|k-1) = \mathbf{A}\mathbf{x}(k-1|k-1)\mathbf{x}(k-1|k-1) + \mathbf{B}\mathbf{u}(k-1) \qquad (13)$$

$$\mathbf{x}(k|k) = \mathbf{x}(k|k-1) + \mathbf{M}_K[\mathbf{y}(k) - \mathbf{C}\mathbf{x}(k|k-1)] \qquad (14)$$

where $\mathbf{M}_K$ is computed via the steady-state solution of the Riccati equation using the Matlab function `kalman` from the covariance matrices $\mathbf{Q}_K = \mathrm{E}\{\mathbf{ww}^T\}$ and $\mathbf{R}_K = \mathrm{E}\{\mathbf{vv}^T\}$, where $\mathbf{w} \in \Re^{N_x}$ and $\mathbf{v} \in \Re^2$ are the assumed white noise disturbance vector signals entering at the system state and output, respectively. However, the KF is used in the sense of an observer, where the diagonal elements of $\mathbf{Q}_K$ and $\mathbf{R}_K$ are used as tuning parameters to achieve desired dynamics.

### 3.1. MPC RWM control simulation results

In the Mathworks Matlab/Simulink development environment, the MPC problem (12) is solved using the modified MPT Toolbox [76] and ILOG/IBM CPLEX as the QP solver. The Simulink block diagram for the proposed MPC ITER RWM control is shown in Figure 3. In this scheme, the original high-order model of plasma RWM dynamics (4-5) is used. The initial state for the simulations was set so that in the modal form (7) the states of the unstable modes were set to a perturbed value of 0.5 while the states of the stable modes were set to 0, unless stated otherwise. Because we want to emphasize performance near actuator voltage constraints, the limit values are set to a relatively low value $|\mathbf{u}| \leq 34$ V.

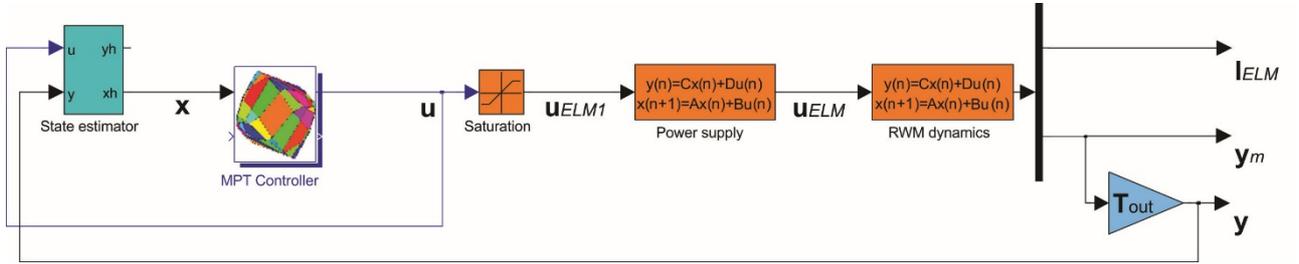

Figure 3. Simulink block diagram for MPC ITER RWM control

With the model (8-9) in the modal form, the control cost $\mathbf{Q}_C$ was tuned so that the two diagonal elements corresponding to the unstable modes were set to 10, the other diagonal elements corresponding to the stable modes were set to $10^{-1}$, while all non-diagonal elements were zero; the cost $\mathbf{R}_C$ was set to $10^{-2} \cdot \mathbf{I}_{27}$. In the KF, the diagonal elements of $\mathbf{Q}_K$ were $10^{-2}$ for the stable modes and $10^{-1}$ for the unstable modes, respectively, and $\mathbf{R}_K$ was $\mathbf{I}_2$. This tuning is a compromise between the requirements for controller responsiveness needed for the stabilisation of the unstable modes and the suppression of measurement noise. The tuning parameters were chosen manually using time-domain simulations and frequency-domain diagrams of the sensitivity functions (see Chapter 7 in [55], and [75]). The latter are a valuable tool because they facilitate the assessment of the robustness to unstructured model uncertainty at high frequencies.

The horizon length $N$ is a specific tuning parameter of MPC. With infinite-horizon MPC, its role is less pronounced than with the traditional finite-horizon form, because the same cost is applied after the end of the finite horizon. However, it affects the length of the horizon in which the constraints are actively considered. With actuator saturations lasting longer than $N$ samples, the constraints are not handled optimally. If $N$ is very small, the horizon in which the constraints are considered is short, and the performance improvement near constraints is restricted. $N$ also affects the dimensions of the resulting QP, and consequently the computational load of the QP solver; high values of $N$ may also result in poor numerical conditioning of the QP problem. The final controller selected value $N = 80$ is shown to be acceptable from the computational aspect; also, higher values do not contribute much more to the practical disturbance suppression ability of the controller. For illustration, in typical simulation experiments (as shown in Figure 4) where the controller brings the system from a perturbed state of the unstable RWM mode to the origin with actuator saturation of the control signal $|\mathbf{u}| \leq 34$ V, the settling times to bring the outputs $\mathbf{y}$ within 0.1 from the origin for $N$ values of 1, 40 and 80, are 0.2012 s, 0.1665 s, and 0.1622 s, respectively. But due to the short sampling time, even with the relatively long horizon $N = 80$ the handling of constraints cannot be considered strictly optimal; the prediction horizon covers 0.06 s, which is less than a half of the actual control saturation period in simulation.

Move blocking [55, 56] is very efficient for reduction of the computational load at longer horizons. Typically, we use blocking of the control signal within the predictive horizon to $N_u = 3$ intervals of lengths (2, 2, 76), resulting in additional equality constraints $\mathbf{u}_0 = \mathbf{u}_1$, $\mathbf{u}_2 = \mathbf{u}_3$, and $\mathbf{u}_4 = \mathbf{u}_5 = ... = \mathbf{u}_{N-1}$. Figure 4 shows the simulation result of the final MPC controller that uses move blocking, which is almost indistinguishable from the simulation with no move blocking; the outputs $\mathbf{y}$ are brought within 0.1 from the origin in 0.1594 s.

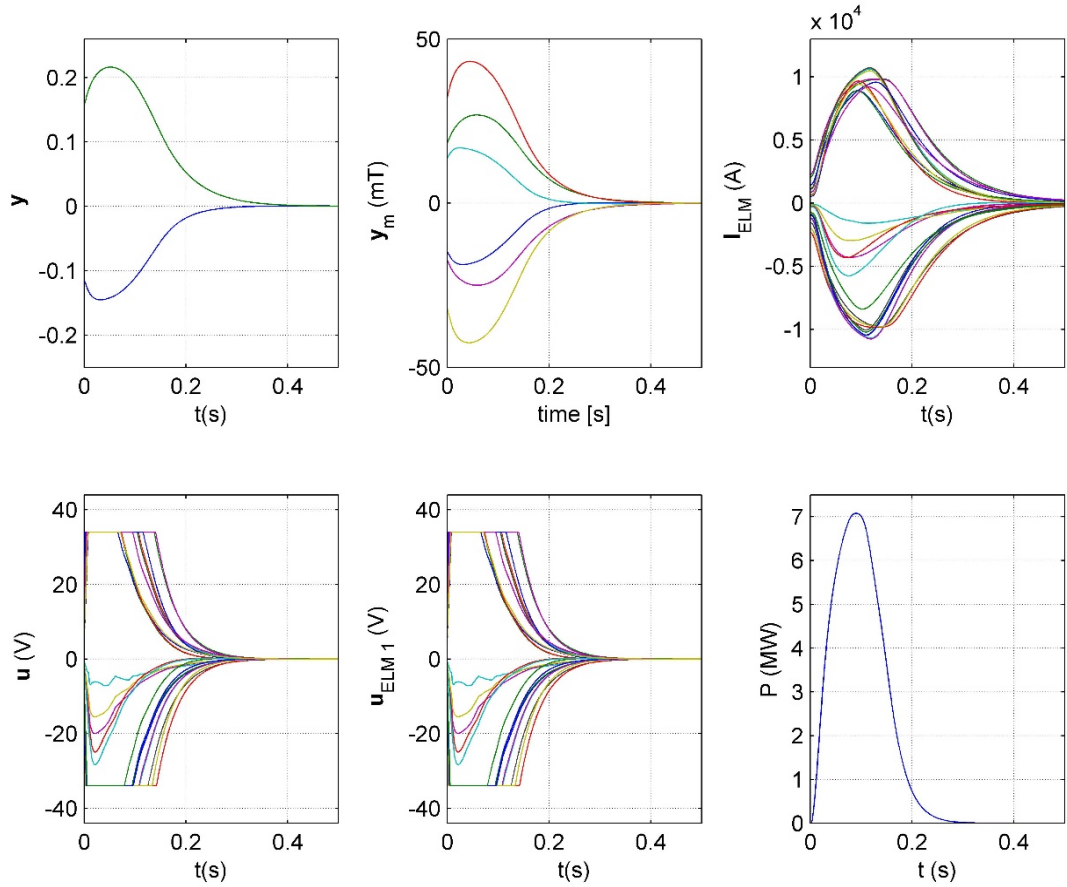

Figure 4. MPC control simulation: $N = 80$, move blocking to 3 intervals of lengths (2, 2, 76), $|\mathbf{u}| \leq 34$ V. Signals: reduced-dimensional measurement vector $\mathbf{y}$ (top-left); measurement output signal $\mathbf{y}_m$ (top-centre); vector of ELM coil currents $\mathbf{I}_{ELM}$ (top-right); control signal vector (PS input) $\mathbf{u}$ (bottom-left); vector of PS voltages after saturation $\mathbf{u}_{ELM1}$ (bottom-centre); power at ELM coils (bottom-right).

Figure 5 shows a simulation of the MPC controller with $N = 80$ and move blocking to 3 intervals of lengths (2, 2, 76) where the control signal constraints are relaxed to $|\mathbf{u}| \leq 144$ V but are not violated during the simulation from the same initial conditions. It is observed that the control signal $\mathbf{u}$ amplitudes are much higher, peaks reaching up to 91 V; the responses of coil currents $\mathbf{I}_{ELM}$ are faster but most of them not reaching higher amplitudes; the outputs $\mathbf{y}$ are brought within 0.1 from the origin notably faster in 0.073 s.

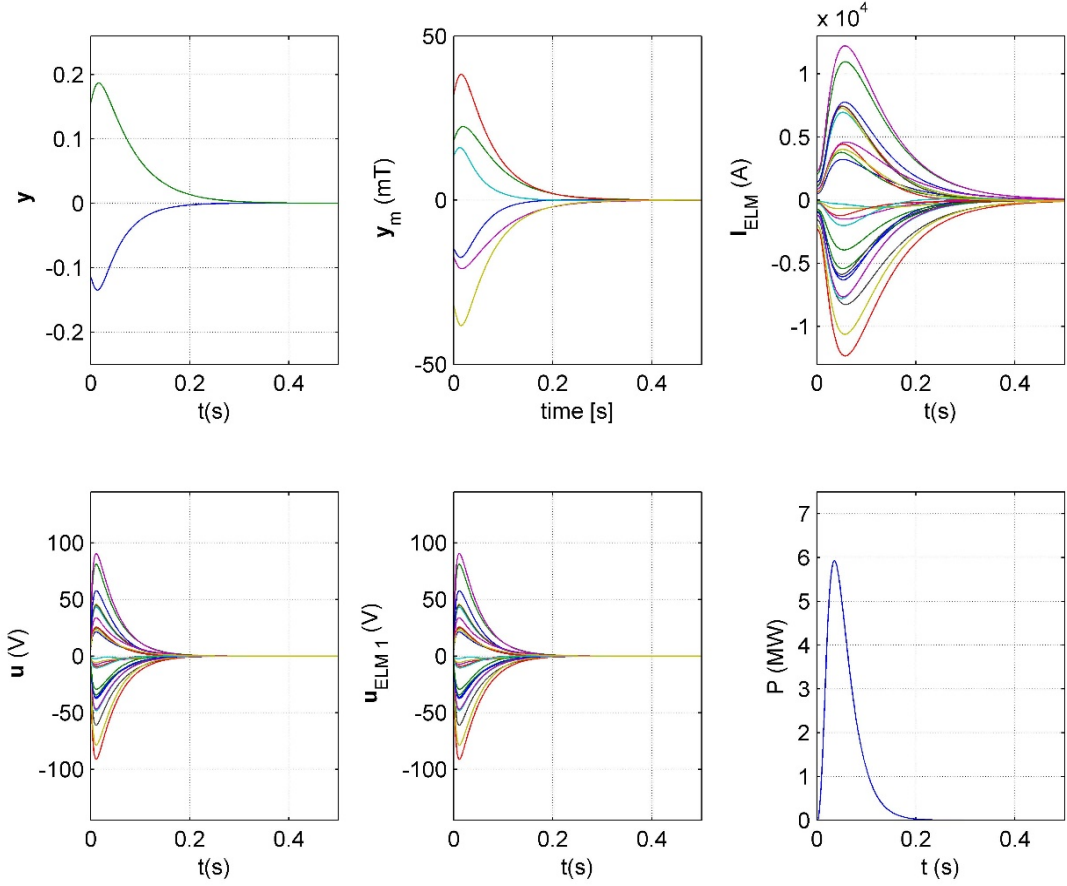

Figure 5. MPC control simulation: $N = 80$, move blocking to 3 intervals of lengths (2, 2, 76), $|\mathbf{u}| \leq 144$ V (the constraints are inactive). Signals as in Figure 4.

### 3.2. LQG control

A performance comparison of MPC control with LQ control is of high interest, as the two controllers share a very similar cost function, and in the absence of constraints and move blocking actually result in equal performance.

The LQ controller is based on the control-oriented model (8) and the cost function

$$J_{\text{LQ}}(k) = \tfrac{1}{2}\sum_{i=0}^{\infty} \mathbf{x}_{k+i}^{T}\mathbf{Q}_{C}\mathbf{x}_{k+i} + \mathbf{u}_{k+i}^{T}\mathbf{R}_{C}\mathbf{u}_{k+i} \qquad (15)$$

and in the commonly used steady-state form the control signal $\mathbf{u}(k)$ is computed as

$$\mathbf{u}(k) = \mathbf{K}_{\text{LQ}}\mathbf{x}(k), \qquad \mathbf{K}_{\text{LQ}} = -(\mathbf{R}_C + \mathbf{B}^T\mathbf{P}\mathbf{B})^{-1}\mathbf{B}^T\mathbf{P}\mathbf{A} \qquad (16)$$

where $\mathbf{P}$ is the steady-state solution of the discrete-time algebraic Riccati equation (11) [74].

The same steady-state Kalman filter (13-14) as with the MPC controller obtains the estimate $\mathbf{x}(k|k)$, that is used in place of the non-measurable system state $\mathbf{x}(k)$.

Differently from Ariola and Pironti [47] where a reduced-dimensional control signal with 6 elements is used, we use the full-dimensional control signal $\mathbf{u}$ with 27 components. The difference observed between the two in the simulation performance is small and is attributed to differences in the model reduction and the tuning parameters.

Figure 6 shows a schematic diagram of the LQG ITER RWM control system, comprising the original high-order model of plasma RWM dynamics (4-5), a model of PS dynamics with saturation, the LQ controller (matrix gain block), and the KF state estimator block. Figure 7 shows the simulation performance of the LQG controller with actuator saturation of the control signal $|\mathbf{u}| \leq 34$ V, which may be compared with MPC controller simulation performance in Figure 4. The LQG controller is able to stabilise the system despite the vicinity of constraints, but its performance is notably worse than that of the MPC controllers, and the outputs $\mathbf{y}$ are brought within 0.1 from the origin in 0.370 s. The LQ controller does not take control signal saturation limits into account, and the control signal after the actuator saturation $\mathbf{u}_{ELM1}$ (bottom-centre) is heavily clipped compared to the LQ controller output signal $\mathbf{u}$ (bottom-left).

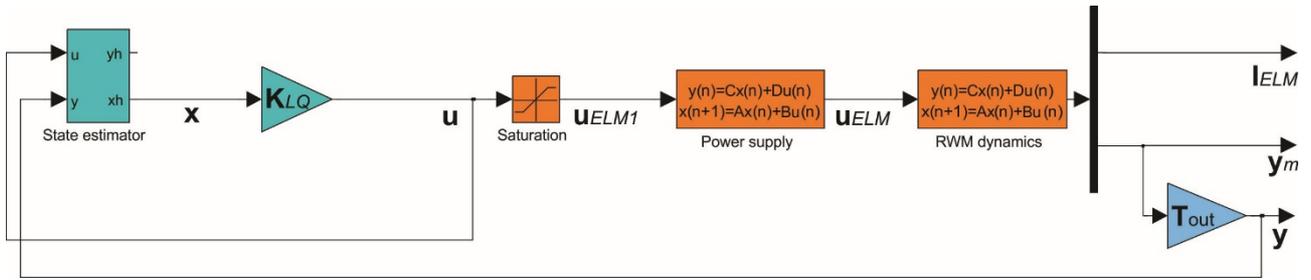

Figure 6. Simulink block diagram for LQG ITER RWM control

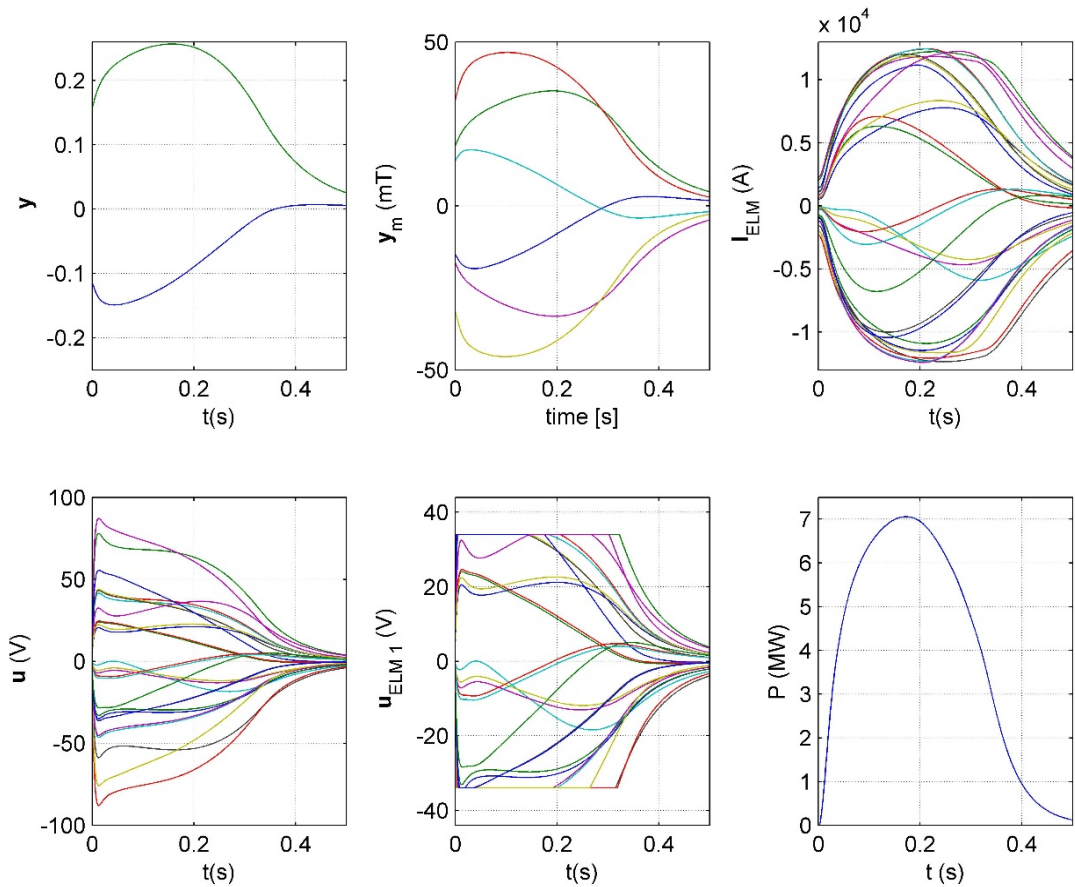

Figure 7. LQG control simulation: $|u| \leq 34$ V. Signals as in Figure 4.

Estimator wind-up protection (EWP) is a relatively simple non-linear modification of the LQG controller in the standard linear form that may be used to improve its performance in the vicinity of actuator constraints. The performance of the standard linear LQG controller is adversely affected by the fact that the KF receives the control signal as computed by the LQ controller, disregarding that it gets clipped at the PS. For EWP, clipping of the control signal with the respective saturation limits is performed with an additional saturation block immediately following the LQ controller block in the modified block diagram in Figure 8, so that the state estimator receives the control signal values actually applied to the plasma. LQG with EWP coincides with our MPC controller (without move blocking) with the degenerate horizon $N = 0$. Figure 9 shows the simulation performance of the LQG controller with EWP; the performance is much better than without EWP and is closer to that of MPC control, and the outputs **y** are brought within 0.1 from the origin in 0.2036 s.

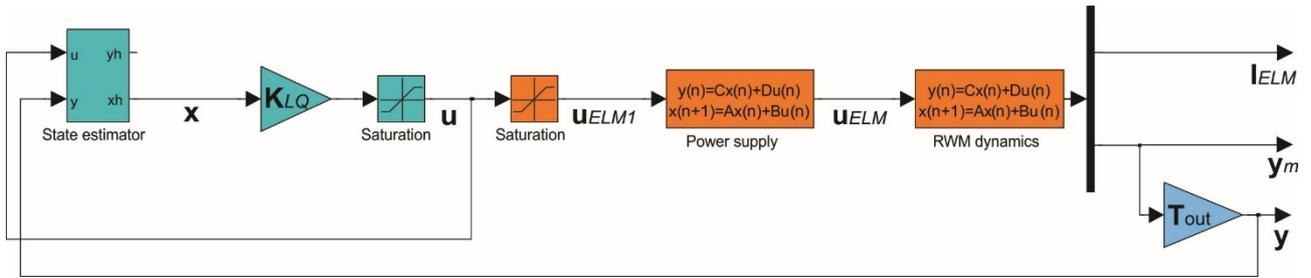

Figure 8. Simulink block diagram for LQG ITER RWM control with estimator wind-up protection

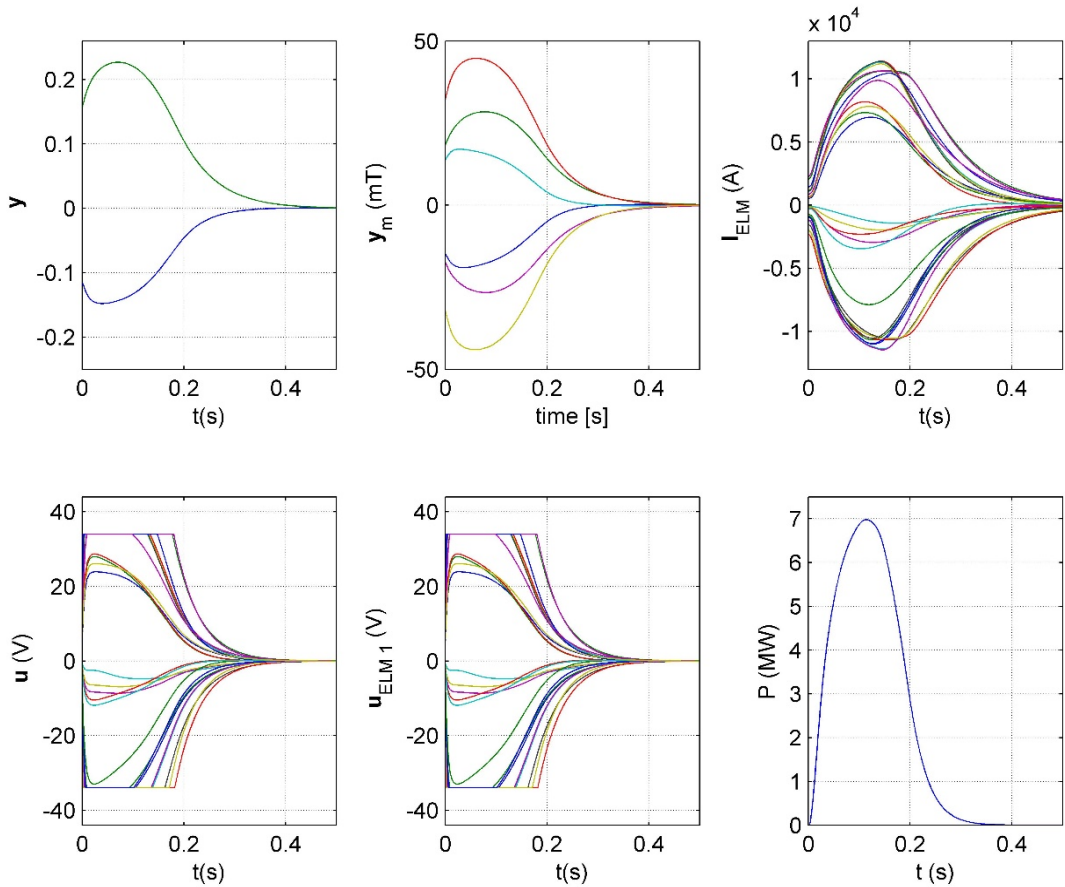

Figure 9. LQG control simulation with estimator wind-up protection: $|\mathbf{u}| \leq 34$ V. Signals as in Figure 4.

### 3.3. Sensitivity to noise

The sensitivity of the closed-loop system with the MPC controller[3] to measurement and actuator noise was assessed in simulation. Figure 10 shows the simulation result where a noise vector generated with a "band-limited white noise" block with noise power $10^{-2}$, at the signal-to-noise (power) ratio (SNR) 13 dB, was injected to the actuator signal. Specifically, it was added to the signal $\mathbf{u}_{ELM1}$ immediately following the Saturation block, as to avoid clipping of the added noise. Despite the relatively high level of noise at $\mathbf{u}_{ELM1}$ (bottom-centre), other displayed signals exhibit very little impact of noise, which is mainly because the noise is heavily suppressed by the plasma process dynamics.

---

[3] In the absence of constraints, the properties of the MPC and LQG controllers are the same. The simulation analysis does not show unexpected differences between them when the constraints are active.

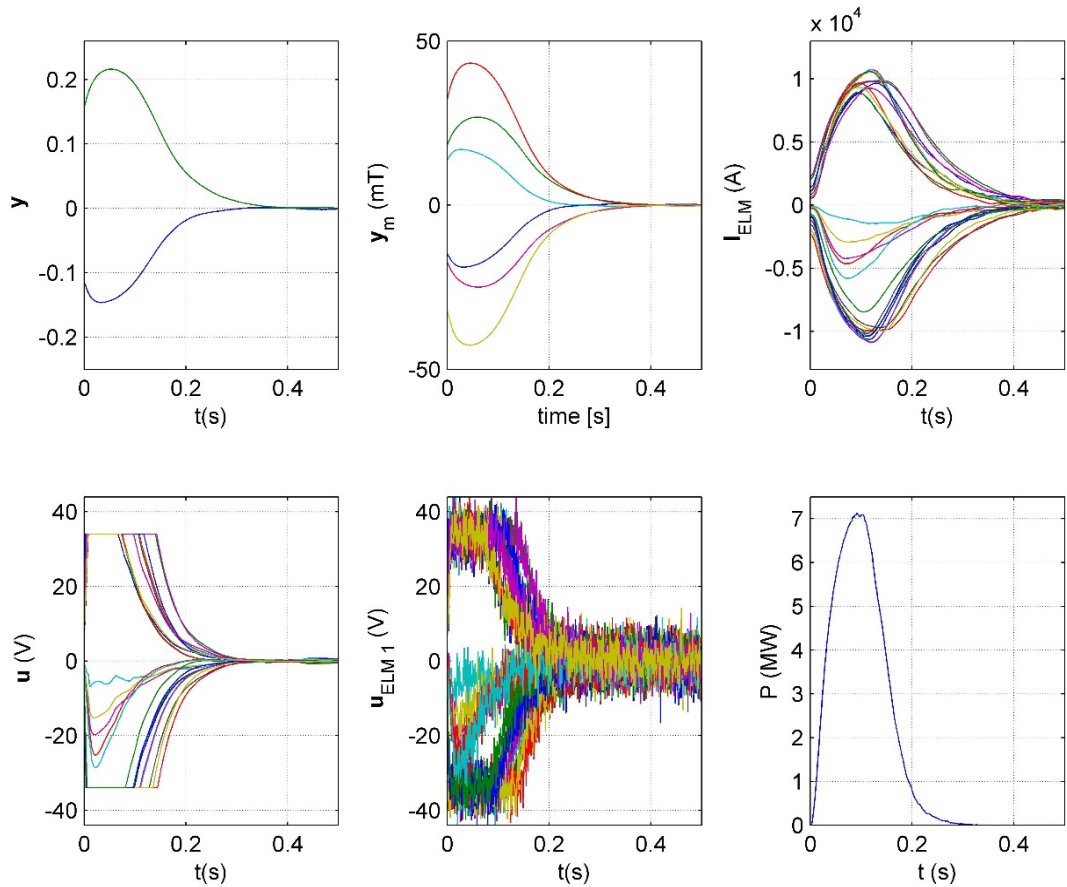

Figure 10. MPC control simulation with actuator noise injected at $\mathbf{u}_{ELM1}$: $N = 80$, move blocking to 3 intervals of lengths (2, 2, 76), $|\mathbf{u}| \leq 34$ V, noise power $10^{-2}$. Signals as in Figure 4.

Figure 11 displays the simulation result where a noise vector generated with a "band-limited white noise" block with noise power $10^{-7}$, with the SNR 18 dB, was injected to the measurements vector $\mathbf{y}_m$. The MPC controller and the KF are tuned for an acceptable level of measurement noise suppression. The noise suppression level may be increased with retuning, but this has an adverse effect upon the efficiency of stabilisation of the unstable RWM dynamics. In case high measurement noise levels are present in practice, they may be reduced by implementing measurements and noise filtering at a higher sampling frequency.

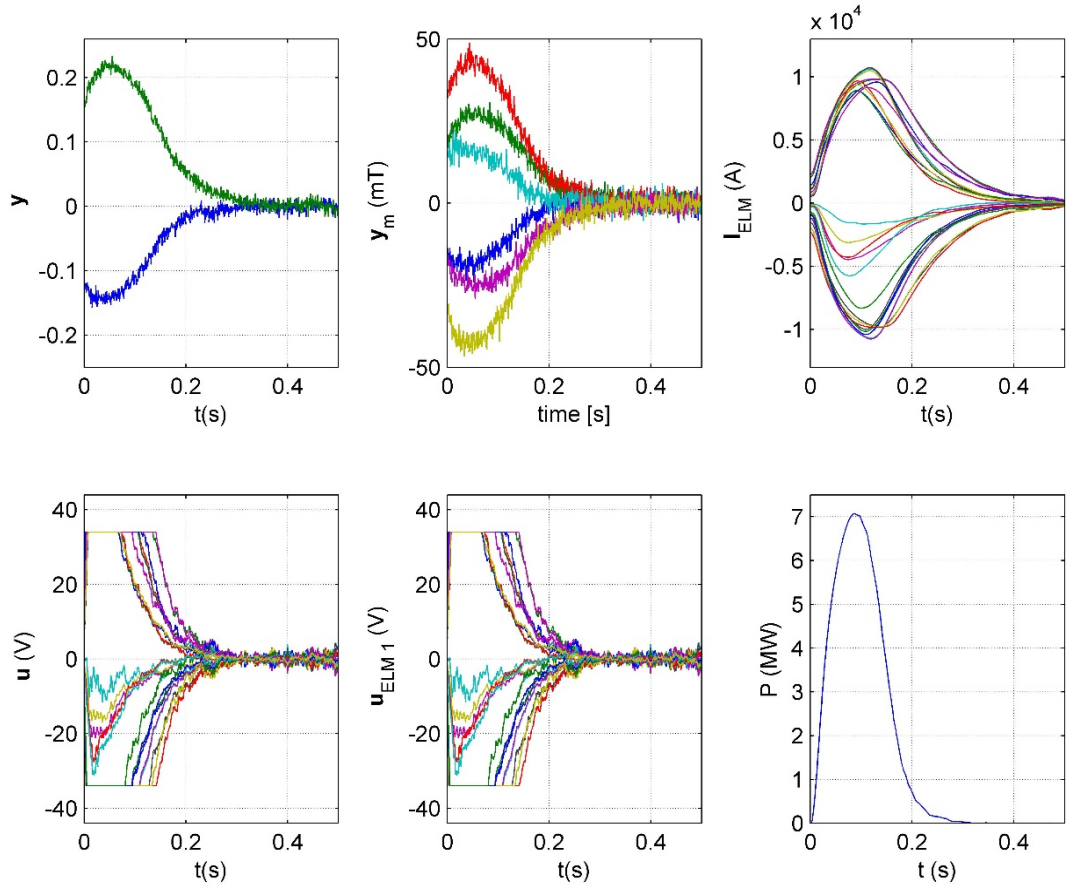

Figure 11. MPC control simulation with measurement noise injected at $\mathbf{y}_m$: $N = 80$, move blocking to 3 intervals of lengths (2, 2, 76), $|\mathbf{u}| \leq 34$ V, noise power $10^{-7}$. Signals as in Figure 4.

### 3.4. Robustness to RWM parameters

Using the state-space transformation of the original RWM model to the modal form and decomposition to the unstable and stable parts (7) we have tested the robustness of the closed-loop performance with the MPC controller[4] (designed using the unmodified model) in simulation with modified values of the growth rate $\gamma$ and the frequency $\omega$.

Figure 12 shows a simulation where $\gamma$ was increased to 20 s$^{-1}$ (from the nominal value of about 19 s$^{-1}$). Even with the small increase of $\gamma$, the system becomes more unstable and the time to bring the outputs $\mathbf{y}$ within 0.1 from the origin increases to 0.2112 s (from 0.1594 s). The reason for this is that the initial point of the simulation is relatively close to the edge of the stabilizable region with the actuator saturation limit $|\mathbf{u}| \leq 34$ V. Increasing $\gamma$ further very soon results in an unstable response. With the actuator saturation limit removed, stable response may be achieved up to 31 s$^{-1}$, with peak values in $\mathbf{u}$ around 105 V.

---

[4] In the absence of constraints, the properties of the MPC and LQG controllers are the same. The simulation analysis does not show unexpected differences between them when the constraints are active.

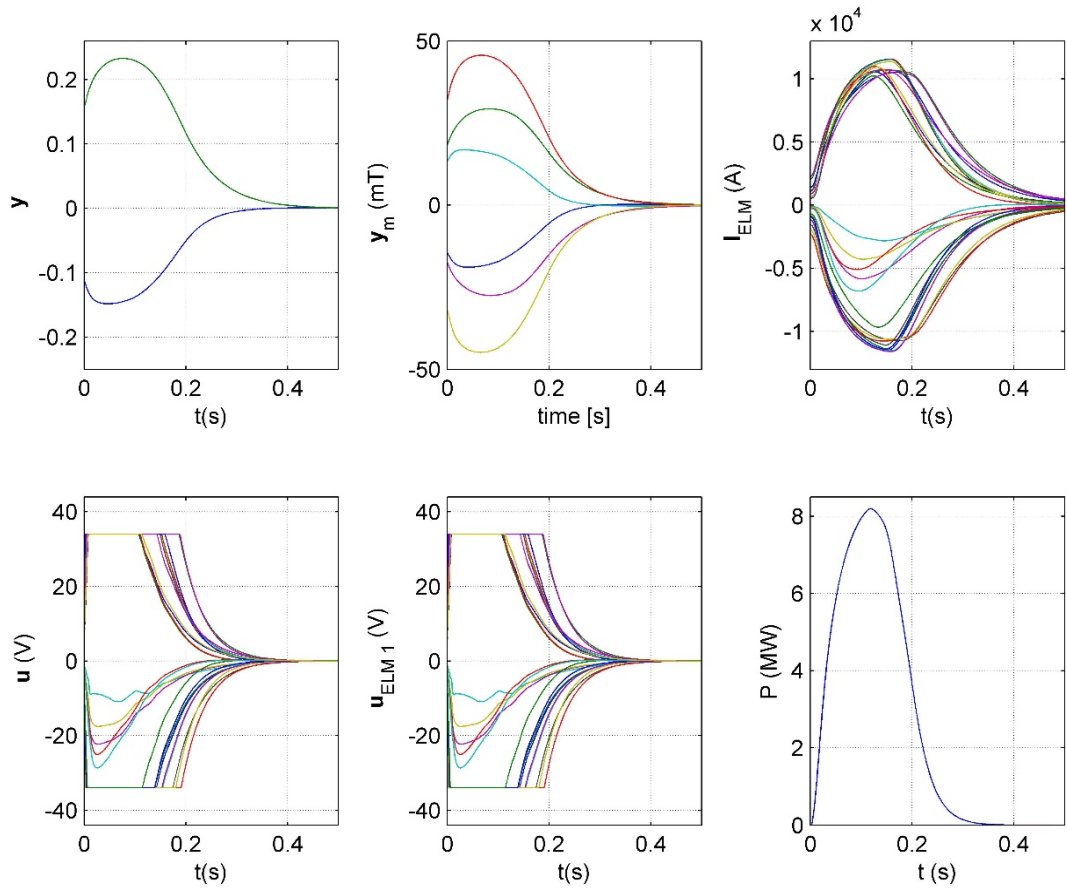

Figure 12. MPC control simulation with increased growth rate $\gamma = 20$ s$^{-1}$: $N = 80$, move blocking to 3 intervals of lengths (2, 2, 76), $|\mathbf{u}| \leq 34$ V. Signals as in Figure 4.

When the growth rate $\gamma$ is decreased, the system dynamics become less unstable. Even at considerably lower and negative (stable) values of $\gamma$, where the nominal model used for designing the MPC controller may be considered highly inaccurate, the observed simulation responses are well-behaved. Figure 13 is an example of such simulation with $\gamma = 0.1$.

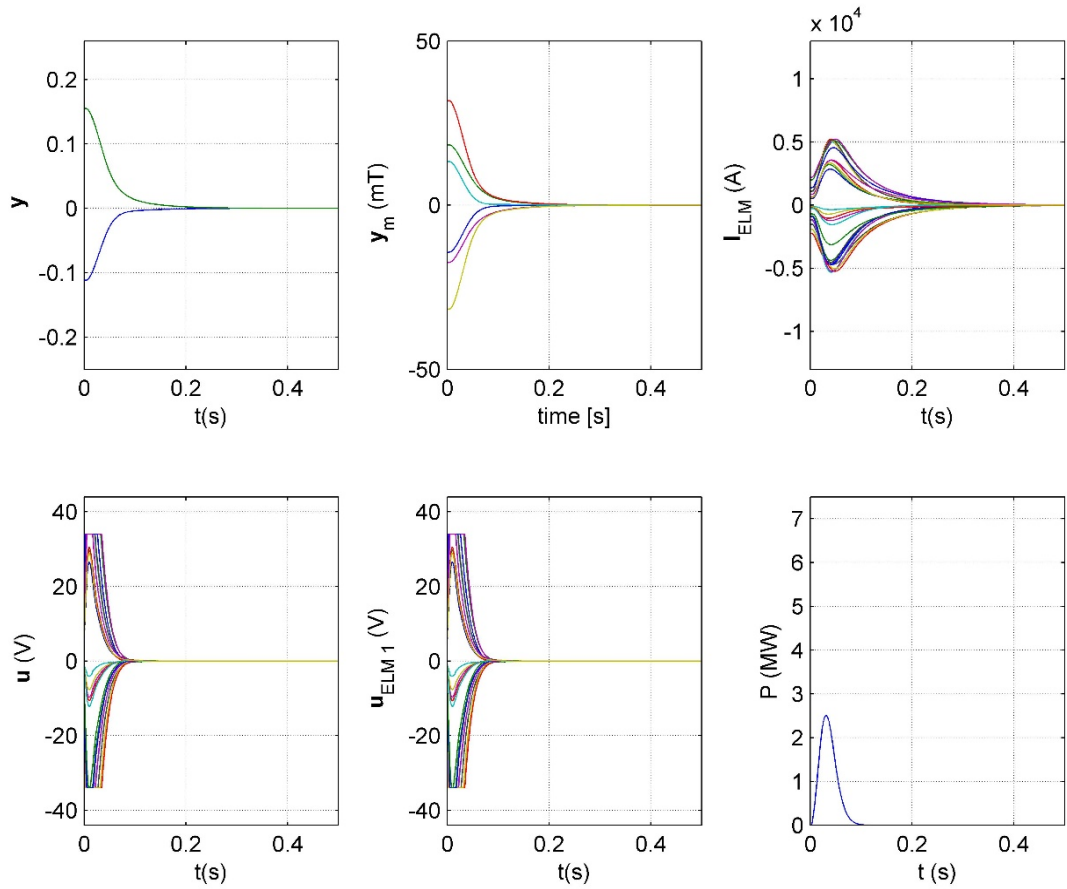

Figure 13. MPC control simulation with increased growth rate $\gamma = 0.1$ s$^{-1}$: $N = 80$, move blocking to 3 intervals of lengths (2, 2, 76), $|\mathbf{u}| \leq 34$ V. Signals as in Figure 4.

When modifying the value of the frequency $\omega$ (nominally about 0.26 rad/s), stable simulation responses were obtained in the range between –24 rad/s and 23 rad/s, possibly exhibiting damped oscillation. Figures 14 and 15 show sample simulated closed-loop responses at $\omega = 15$ rad/s and $\omega = -15$ rad/s, respectively.

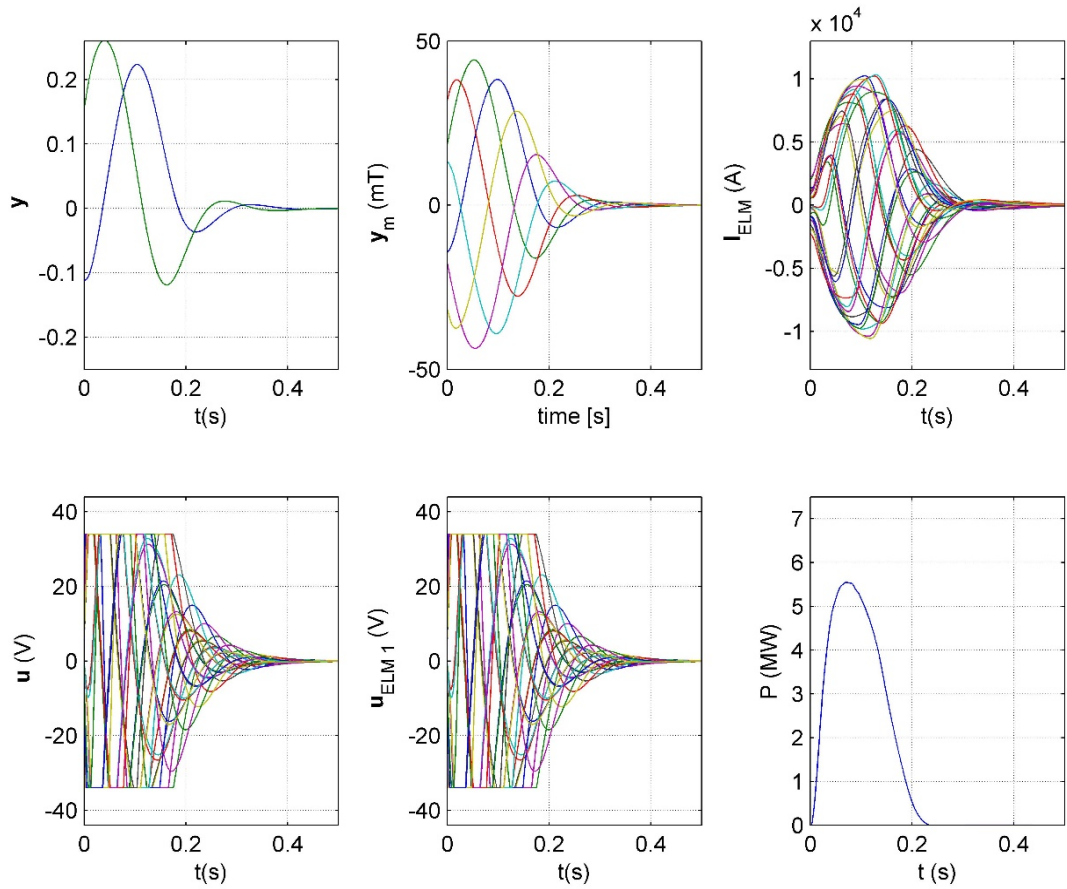

Figure 14. MPC control simulation with increased frequency $\omega = 15$ rad/s: $N = 80$, move blocking to 3 intervals of lengths (2, 2, 76), $|\mathbf{u}| \leq 34$ V. Signals as in Figure 4.

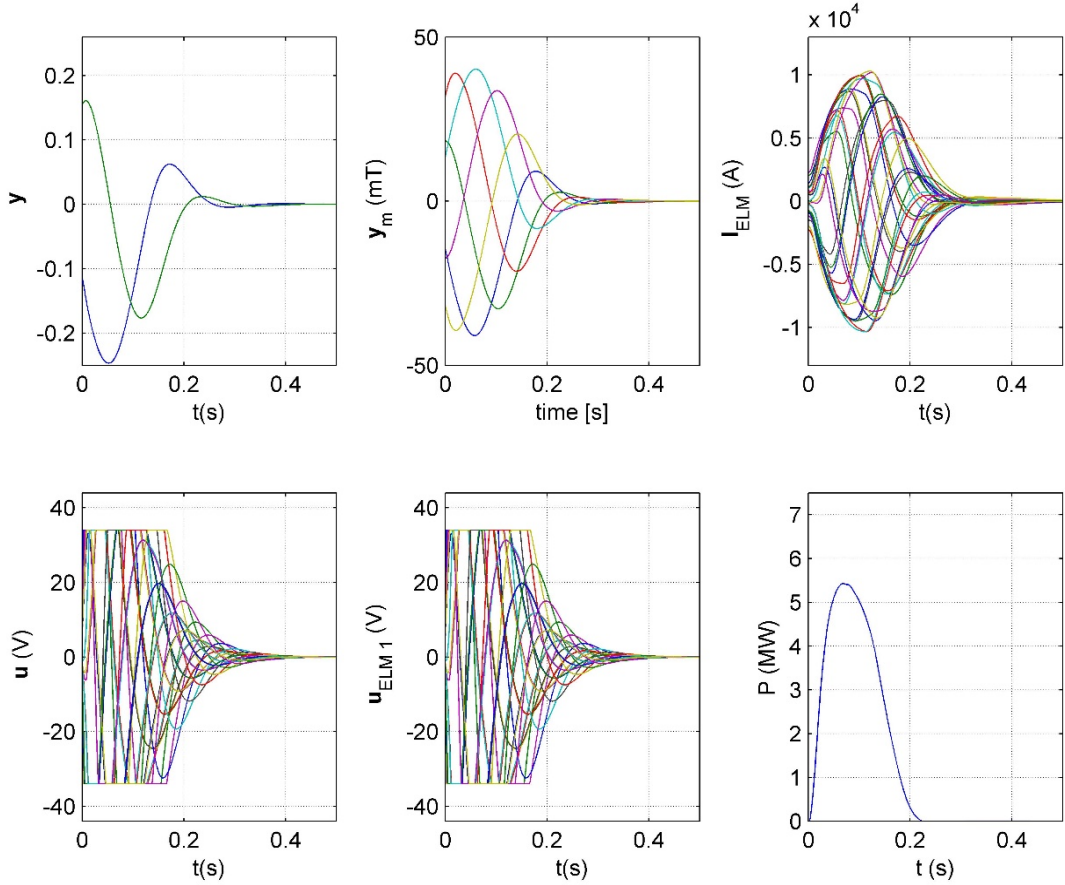

Figure 15. MPC control simulation with increased frequency $\omega = -15$ rad/s: $N = 80$, move blocking to 3 intervals of lengths (2, 2, 76), $|\mathbf{u}| \leq 34$ V. Signals as in Figure 4.

### 3.5. Region of stability

Ariola and Pironti [47] evaluated the null-controllable region of the ITER RWM LQ controller considering only the unstable system dynamics subject to coil voltage constraints, mapped it to the best achievable performance (BAP) in terms of initial maximum measured initial magnetic field displacements, and compared it with the achieved simulation results of the LQG RWM controller. Theoretical evaluation of the stabilizable region of the state-space is also possible for infinite-horizon MPC controllers with the LQ terminal cost and the constraint to its associated feasible terminal set [57] for small-dimensional systems; it is known that the domain of attraction grows with increasing horizon $N$.

We are most interested in practically comparing the BAP of the closed-loop system with our MPC controller and the KF with that of the system with the closely related LQG controller with EWP, which were evaluated in simulation with the full-dimensional RWM model with the PS saturation $|\mathbf{u}| \leq 34$ V. Using the state-space transformation of the original RWM model to the modal form and decomposition to the unstable and stable parts (7) we explored BAP in terms of the stabilizable region of the two unstable modes, $\xi_1$ and $\xi_2$. In Figure 16, the height represents the integral of PS power until the simulation termination at 0.5 s (for stable closed-loop responses), at grid points of an edge section of the region of the unstable dynamic states, $0.35 \leq \xi_1 \leq 0.65$ and $0.35 \leq \xi_2 \leq 0.65$. Unstable and borderline responses where the power integral exceeds 5 MWs are omitted. It can be observed that in the central area of the stabilizable region the difference between the two controllers in the value of the power integral is very small, but towards the edge the MPC controller requires less power and finally provides a 7% larger stabilizable area.

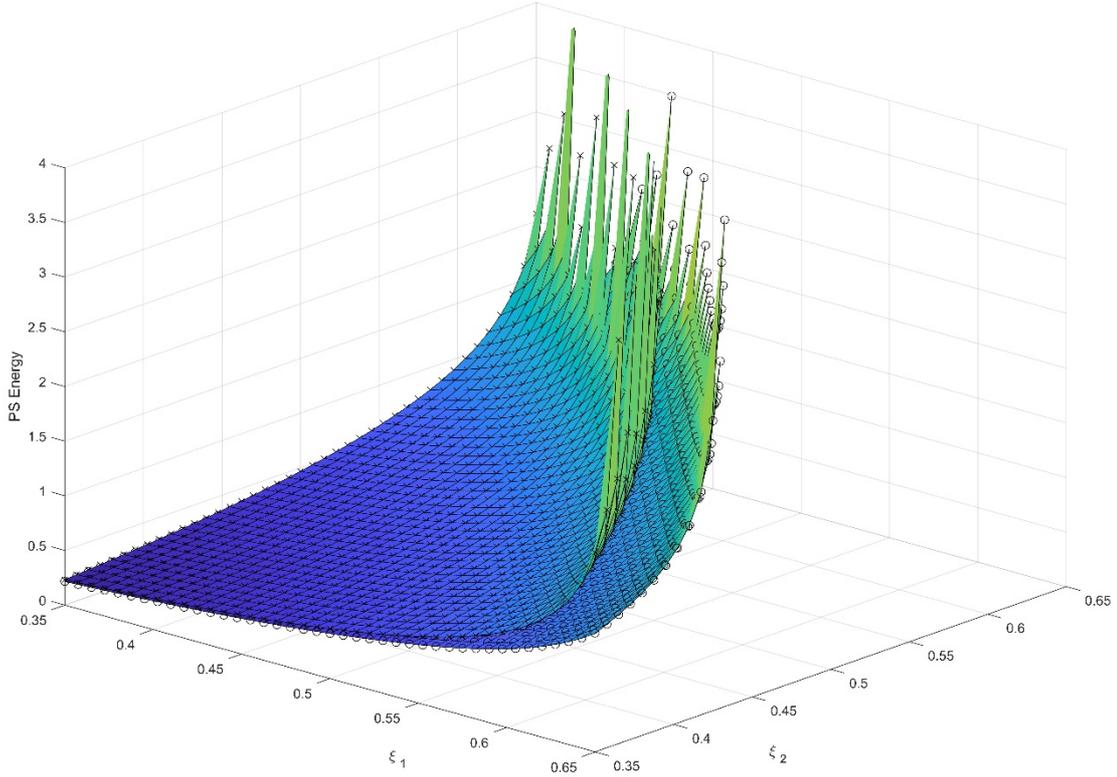

Figure 16. Best achievable performance for actuator saturation $|\mathbf{u}| \leq 34$ V in terms of the stabilizable region of the unstable modes $\xi_1$ and $\xi_2$ (only an edge section of the region is shown). The height at each pair of unstable mode values represent the integral of PS power in simulation until 0.5 s if the response is stable (×: LQG control with EWP; ○: MPC control). MPC: $N = 80$, move blocking to 3 intervals of lengths (2, 2, 76).

It should be pointed out that the proposed MPC scheme is not able to handle state constraints, such as constraints on the coil currents, and that the coil currents in the shown simulations exhibit coil currents consuming a considerable portion of the maximum coil current capacity 15 kA, which will likely have to be shared with the primary ELM coil function. A programmable saturation block may be inserted into the control scheme to limit the range of the ELM coil current available for RWM control. However, the suppression of the unstable RWM requires a sufficient counter-acting corrective magnetic field, for which adequate ELM coil currents are needed. Even if a computationally feasible MPC implementation that would also consider coil current constraints is found, it cannot be expected to reduce the coil currents arbitrarily. Hence, in practical experiment, the size of the stabilizable region of RWM perturbations will depend on the ELM voltage and current ranges assigned to RWM control.

## 4. Solving QP using primal FGM

The C-language implementation of the QP solver is based on the generalized primal FGM algorithm [72] of the QPgen code generator [63]. This is a relatively simple optimization algorithm, known to be very efficient for MPC control problems with input constraints. Compared to second-order methods, the convergence rate is not distinguished, but FGM iterations are simple, and not many are required typically for the precision demanded for control.

FGM belongs to the family of first-order proximal optimization methods, designed for problems of the minimization of a sum of two functions with specific properties. In our case, we are solving the problem

$$\min\bigl(J(\tilde{\mathbf{u}}) + \psi(\tilde{\mathbf{u}})\bigr) \qquad (17)$$

where

- the optimization variable is the vector of the controller outputs over the predictive horizon, $\tilde{\mathbf{u}} = [\mathbf{u}_k, \ldots, \mathbf{u}_{k+N-1}]$ (using the *condensed* approach, the system state **x** does not appear in the optimization variable, because the equality constraints of the control model state transition equation (8) are directly substituted into the MPC cost function (10); using move blocking, the dimension of $\tilde{\mathbf{u}}$ is reduced to $27 \cdot N_u$);

- the first function $J$ is the MPC cost function (10), as a quadratic function of $\tilde{\mathbf{u}}$ known to be strongly convex with known convexity parameter, differentiable, and with gradient satisfying the generalized Lipschitz property

$$J(\tilde{\mathbf{u}}_1) \leq J(\tilde{\mathbf{u}}_2) + (\tilde{\mathbf{u}}_1 - \tilde{\mathbf{u}}_2)^T \nabla J(\tilde{\mathbf{u}}_2) + \frac{1}{2} \|\tilde{\mathbf{u}}_1 - \tilde{\mathbf{u}}_2\|_\mathbf{L}^2 \qquad (18)$$

for all $\tilde{\mathbf{u}}_1, \tilde{\mathbf{u}}_2 \in \mathbb{R}^{27N_u}$, where the $\mathbf{L}$-norm is $\|\mathbf{x}\|_\mathbf{L} = \sqrt{\mathbf{x}^T \mathbf{L} \mathbf{x}}$, and $\mathbf{L}$ is a strongly positive definite diagonal matrix replacing a scalar Lipschitz coefficient;

- the second function $\psi$ is the indicator function of the feasible set of the input constraints – proper, closed and convex, defined as

$$\psi(\tilde{\mathbf{u}}) = \begin{cases} 0 & \text{if } \tilde{\mathbf{u}}_{\min} \leq \tilde{\mathbf{u}} \leq \tilde{\mathbf{u}}_{\max} \\ \infty & \text{otherwise} \end{cases} \qquad (19)$$

The generalized proximity operator is generally defined as

$$\operatorname{prox}_\psi^\mathbf{L}(\mathbf{x}) \triangleq \arg\min_\mathbf{y} \left( \psi(\mathbf{y}) + \frac{1}{2} \|\mathbf{y} - \mathbf{x}\|_\mathbf{L}^2 \right) \qquad (20)$$

and with the choice of the indicator function $\psi$ represents a simple projection onto a positive orthant, in practice easily implemented by clipping each vector element with its respective constraint value. $\mathbf{L}$ is determined by solving an optimization problem seeking the optimal preconditioner with diagonal structure (preserving simple projection) which minimises the condition number of the preconditioned QP problem [62, 63].

The QP being solved in the condensed form is stated as

$$\min_{\tilde{\mathbf{u}}} \frac{1}{2} \tilde{\mathbf{u}}^T \mathbf{H}_c \tilde{\mathbf{u}} + \mathbf{f}_c^T \tilde{\mathbf{u}} + c_c$$

$$\text{subject to } \begin{bmatrix} \mathbf{I} \\ -\mathbf{I} \end{bmatrix} \tilde{\mathbf{u}} \leq \begin{bmatrix} \tilde{\mathbf{u}}_{\max} \\ \tilde{\mathbf{u}}_{\min} \end{bmatrix}, \qquad (21)$$

where $\mathbf{H}_c$, $\mathbf{f}_c$, and $c_c$ are the second-order, first-order and constant cost function terms, respectively, and $c_c$ may be omitted as it does not affect the optimization result.

The FGM algorithm is summarized in Algorithm 1 below, where the top indices represent the iterations. It requires a specific scalar sequence $\beta^i$, which may be computed in advance, required to achieve the desired convergence properties. The implementation can be made division-free. Some initialization tasks are required in each time step, for instance the part of the computation of the first-order term vector $\mathbf{f}_c$, which depends on the current system state estimate $\mathbf{x}_k$.

Algorithm 1: Fast gradient method

Initialize: $\mathbf{v}^1 = \tilde{\mathbf{u}}^0 \in \mathbb{R}^{27N_u}$, sequence $\beta^i$, vector $\mathbf{f}_c$

**for** $i = 1$ to $i_{\max}$

 $\mathbf{x}^i = \mathbf{v}^i - \mathbf{L}^{-1}(\mathbf{H}_c \mathbf{v}^i + \mathbf{f}_c)$   (gradient step)

 $\tilde{\mathbf{u}}^i = \operatorname{prox}_\psi^\mathbf{L}(\mathbf{x}^i)$   (projection onto feasible set)

 $\mathbf{v}^{i+1} = \tilde{\mathbf{u}}^i + \beta^i (\tilde{\mathbf{u}}^i - \tilde{\mathbf{u}}^{i-1})$   (acceleration)

 **if** $(\mathbf{v}^i - \tilde{\mathbf{u}}^i)^T (\tilde{\mathbf{u}}^i - \tilde{\mathbf{u}}^{i-1}) > 0$   (adaptive restart)

  $\mathbf{v}^{i+1} = \tilde{\mathbf{u}}^{i-1}, \quad \tilde{\mathbf{u}}^i = \tilde{\mathbf{u}}^{i-1}$

 **endif**

**endfor**

The QP terms are derived from the MPC optimization problem (12) by stacking the equations of the equality constraints as follows

$$\tilde{\mathbf{x}} = \mathcal{A}\mathbf{x}_k + \mathcal{B}\tilde{\mathbf{u}}, \qquad \tilde{\mathbf{x}} = \begin{bmatrix} \mathbf{x}_k \\ \mathbf{x}_{k+1} \\ \mathbf{x}_{k+2} \\ \vdots \\ \mathbf{x}_{k+N} \end{bmatrix}, \qquad \mathcal{A} = \begin{bmatrix} \mathbf{I} \\ \mathbf{A} \\ \mathbf{A}^2 \\ \vdots \\ \mathbf{A}^N \end{bmatrix}, \qquad \mathcal{B} = \begin{bmatrix} \mathbf{0} & \mathbf{0} & \mathbf{0} & \cdots & \mathbf{0} \\ \mathbf{B} & \mathbf{0} & \mathbf{0} & & \mathbf{0} \\ \mathbf{AB} & \mathbf{B} & \mathbf{0} & & \mathbf{0} \\ \vdots & & & & \vdots \\ \mathbf{A}^{N-1}\mathbf{B} & \mathbf{A}^{N-2}\mathbf{B} & \mathbf{A}^{N-3}\mathbf{B} & \cdots & \mathbf{B} \end{bmatrix} \qquad (22)$$

and substituting $\tilde{\mathbf{x}}$ in the MPC cost function (10) rewritten as

$$J(k) = \tfrac{1}{2}\tilde{\mathbf{x}}^T \mathcal{Q}\tilde{\mathbf{x}} + \tfrac{1}{2}\tilde{\mathbf{u}}^T \mathcal{R}\tilde{\mathbf{u}}, \qquad \mathcal{Q} = \begin{bmatrix} \mathbf{Q}_C & \mathbf{0} & \cdots & \mathbf{0} & \mathbf{0} \\ \mathbf{0} & \mathbf{Q}_C & & \mathbf{0} & \mathbf{0} \\ \vdots & & \ddots & & \vdots \\ \mathbf{0} & \mathbf{0} & & \mathbf{Q}_C & \mathbf{0} \\ \mathbf{0} & \mathbf{0} & \cdots & \mathbf{0} & \mathbf{P} \end{bmatrix}, \qquad \mathcal{R} = \begin{bmatrix} \mathbf{R}_C & \mathbf{0} & \cdots & \mathbf{0} \\ \mathbf{0} & \mathbf{R}_C & & \mathbf{0} \\ \vdots & & \ddots & \vdots \\ \mathbf{0} & \mathbf{0} & \cdots & \mathbf{R}_C \end{bmatrix} \qquad (23)$$

yielding

$$\mathbf{H}_c = \mathcal{B}^T \mathcal{Q}\mathcal{B} + \mathcal{R}, \qquad \mathbf{f}_c^T = \mathbf{x}_k^T \mathcal{A}^T \mathcal{Q}\mathcal{B}, \qquad c_c = \tfrac{1}{2} \mathbf{x}_k^T \mathcal{A}^T \mathcal{Q}\mathcal{A}\mathbf{x}_k \qquad (24)$$

### 4.1. FGM solver numerical performance

The results of the generalized primal FGM approach were found to match those obtained in Section 3.1 using the MPT Toolbox [76] with Ilog/IBM CPLEX as the reference QP solver with good accuracy, even with relatively small numbers of iterations.

The accuracy of the FGM QP solver solution after $i$ iterations $\tilde{\mathbf{u}}^i$ is compared to the optimal solution $\tilde{\mathbf{u}}^*$ (in practice this reference solution was computed using the CPLEX solver) with the normalised mean-square-error (MSE) formula

$$MSE(\tilde{\mathbf{u}}^i, \tilde{\mathbf{u}}^*) = \sqrt{\frac{1}{27N_u} \sum_{k=1}^{27N_u} \left( \frac{\tilde{u}_k^i - \tilde{u}_k^*}{\tilde{u}_{\max,k} - \tilde{u}_{\min,k}} \right)^2} \qquad (25)$$

The FGM algorithm was initialised with cold start from zero values. As shown in Figure 17, a sufficiently low worst-case MSE below $10^{-4}$ among all samples of the simulation (as shown in Figure 4) was achieved with $i = 20$ iterations of the FGM algorithm, at cost function deviation $J(\tilde{\mathbf{u}}^i) - J(\tilde{\mathbf{u}}^*)$ below $3\cdot 10^{-4}$. With a strongly-convex objective function, which applies to the quadratic cost function of MPC, the FGM algorithm has a linear convergence rate [72, 62] and would under these terms require 50 iterations for a certified solution. In the presence of constraints this is a non-smooth optimisation problem, but Nesterov [72] proves that with restarting linear convergence extends to the strongly-convex proximal case with input constraints. Recently, a proof of linear convergence for a restarting FGM scheme was outlined also for non-smooth strongly-convex functions [77].

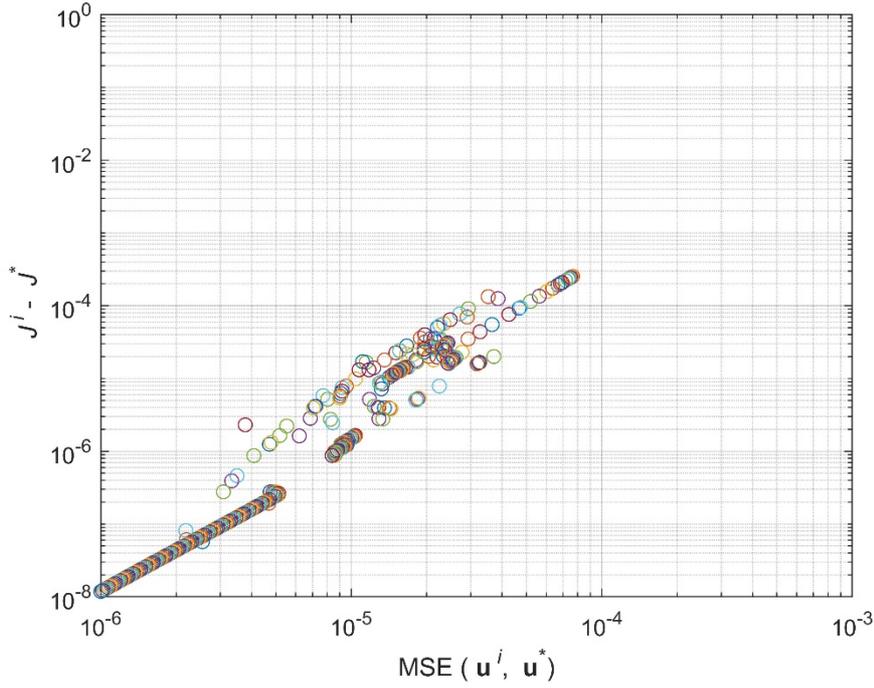

Figure 17. FGM QP solver convergence towards the reference solution after $i$ = 20 iterations. Each circle represent the accuracy of the QP result of one sample of the RWM MPC control simulation ($N$ = 80, move blocking to 3 intervals of lengths (2, 2, 76), $|\mathbf{u}| \leq 34$ V). Horizontal axis: the MSE (25) difference between $\tilde{\mathbf{u}}^{20}$ and $\tilde{\mathbf{u}}^*$. Vertical axis: $J(\tilde{\mathbf{u}}^{20}) - J(\tilde{\mathbf{u}}^*)$.

## 4.2. FGM solver low-latency implementation

In order to test real-time implementability of the MPC RWM controller, the C code generated by QPgen [63] was adapted for fast execution and compiled using Intel C++ Compiler either as a Matlab mex function or in low-latency Linux environment of Ubuntu Linux 14.04, 4.4.9-rt17 SMP PREEMPT RT kernel on a standard desktop computer based on the Intel Core i7-2600K processor (3.4 GHz, 8MB cache, 8 GB RAM). For the execution time measurements, the MPC controller task was running in shielded configuration on dedicated processor cores (standard multi-threading is not useful for accelerating the execution at such short time scales).

Table 1 summarizes the results of the maximum and average computation time and the accuracy of the solution in terms of the MSE and the cost function deviation from the reference solution depending on the numerical precision (double or single, both floating point) and the number of iterations. With double precision, the solver requires on average 26 μs for the prologue and epilogue, and 2.6 μs for each iteration of the main loop. Switching to single precision reduces the computation time by about 20%. The maximum computation time values are about 20% higher than the average ones. Using single precision, the desired worst-case MSE below $10^{-4}$ is reached at maximum computation 0.08 ms, which is considered acceptable with regard to the sampling time 0.75 ms. With single precision, the computation of the cost function deviation $J(\tilde{\mathbf{u}}^i) - J(\tilde{\mathbf{u}}^*)$ is notably sensitive to noise, however the cost function value is not used within the algorithm.

Table 1. FGM solver computation time and worst-case accuracy in RWM MPC over the control simulation ($N = 80$, move blocking to 3 intervals of lengths (2, 2, 76), $|\mathbf{u}| \leq 34$ V), with different floating-point precision and number of iterations

| Precision | No. of iterations $i$ | Max. time (ms) | Av. time (ms) | $\max(MSE(\widetilde{\mathbf{u}}^i, \widetilde{\mathbf{u}}^*))$ | $\max\left(J(\widetilde{\mathbf{u}}^i) - J(\widetilde{\mathbf{u}}^*)\right)$ |
|---|---|---|---|---|---|
| Double | 10 | 0.0743 | 0.0525 | $6.2660 \cdot 10^{-3}$ | $1.6790 \cdot 10^{-0}$ |
| Double | 20 | 0.1052 | 0.0822 | $7.6485 \cdot 10^{-5}$ | $2.5361 \cdot 10^{-4}$ |
| Double | 30 | 0.1345 | 0.1146 | $1.3998 \cdot 10^{-6}$ | $7.8915 \cdot 10^{-8}$ |
| Double | 50 | 0.2093 | 0.1701 | $1.2481 \cdot 10^{-9}$ | $5.8208 \cdot 10^{-11}$ |
| Double | 100 | 0.3326 | 0.2889 | $2.3944 \cdot 10^{-15}$ | $5.8208 \cdot 10^{-11}$ |
| Single | 10 | 0.0691 | 0.0404 | $6.2660 \cdot 10^{-3}$ | $1.6796 \cdot 10^{-0}$ |
| Single | 20 | 0.0802 | 0.0647 | $7.6495 \cdot 10^{-5}$ | $8.0717 \cdot 10^{-4}$ |
| Single | 30 | 0.1063 | 0.0879 | $1.5723 \cdot 10^{-6}$ | $5.5342 \cdot 10^{-4}$ |
| Single | 50 | 0.1458 | 0.1304 | $6.7532 \cdot 10^{-7}$ | $5.5341 \cdot 10^{-4}$ |
| Single | 100 | 0.2553 | 0.2361 | $6.6801 \cdot 10^{-7}$ | $5.5341 \cdot 10^{-4}$ |

## 5. Conclusions

An infinite-horizon MPC controller for active feedback stabilization of the RWM for an ITER equilibrium above the no-wall $\beta_N$ limit in the presence of input (ELM coil voltage) constraints is presented and compared in simulation with closely related LQG control. As expected, the performance of the two is very similar in the absence of constraints. When coil voltage constraints do become active, close to the edge of the stabilizable region of the initial values of the unstable modes, the MPC controller is able to stabilize the system faster and with less control power, and increases the stabilizable region of the RWM perturbation. The advantage of MPC increases at longer horizon $N$; the move-blocking technique is efficient for the reduction of the associated computation demand. A significant part of the performance improvement, compared to the textbook linear LQG, can be achieved already with the small EWP modification of LQG, which may be interpreted as degenerate MPC with $N = 0$. For the simulated conditions, with MPC the stabilizable region is increased by 7% compared to MPC with EWP. In simulations, the sensitivity to noise is reasonable, and the closed-loop system is shown to be robust to changes of unstable mode dynamics. Despite being designed for a single nominal model, the simulated responses are well-behaved over a wide range of growth rates (including stable RWMs) and frequencies.

The method of solving the QP associated with the MPC on-line optimization problem using primal FGM is also described. The on-line computational demand of this QP solver is over an order of magnitude higher than that of the matrix-to-vector multiplications needed for the LQG controller. However, the implementation using a standard desktop computer is sufficiently fast for application in ITER; and a much faster FPGA implementation suitable for experimentation on smaller, dynamically faster tokamaks is known to be feasible.

This MPC implementation cannot optimize the performance with respect to ELM coil current constraints. The latter may be addressed using dual FGM, however it would be computationally much more demanding. Also, a substantial reduction of the coil currents cannot be expected, because an adequate corrective magnetic field is essential for the RWM stabilization. In practical conditions, the stabilizable range of the RWM perturbations will depend on the portion of the current and voltage ranges of the ELM coils assigned to RWM control.

## Funding

Support by Slovenian Research Agency (P2-0001) is gratefully acknowledged.

## References

[1] Walker M. L., Humphreys D. A., Mazon D., Moreau D., Okabayashi M, Osborne T. H., Schuster E., "Emerging applications in tokamak plasma control", IEEE Control Systems Magazine **25** (2005). DOI: 10.1109/MCS.2005.1512797
[2] Chu M. S., Okabayashi M., "Stabilization of the external kink and the resistive wall mode", Plasma Physics and Controled Fusion **52** (2010). DOI: 10.1088/0741-3335/52/12/123001
[3] Strait E. J., "Magnetic control of magnetohydrodynamic instabilities in tokamaks", Physics of Plasmas **22** (2015). DOI: 10.1063/1.4902126
[4] Strait E. J., Bialek J., Bogatu N., Chance M., Chu M. S., Edgell D., Garofalo A. M., Jackson G. L., Jensen T. H., Johnson L. C., Kim J. S., La Haye R. J., Navratil G., Okabayashi M., Reimerdes H., Scoville J. T., Turnbull A. D., Walker M. L., the DIII–D Team, "Resistive wall stabilization of high-beta plasmas in DIII–D", Nuclear Fusion **43** (2003). DOI: 10.1088/0029-5515/43/6/306
[5] Liu Y., Bondeson A., Gribov Y., Polevoi A. "Stabilization of resistive wall modes in ITER by active feedback and toroidal rotation", Nuclear Fusion **44** (2004). DOI: 10.1088/0029-5515/44/2/003

[6] Matsunaga G., Aiba N., Shinohara K., Sakamoto Y., Isayama A., Takechi M., Suzuki T., Oyama N., Asakura N., Kamada Y., Ozeki T. (JT-60 Team), "Observation of an Energetic-Particle-Driven Instability in the Wall-Stabilized High-β Plasmas in the JT-60U Tokamak", Physical Review Letters **103** (2009). DOI: 10.1103/PhysRevLett.103.045001

[7] Berkery J. W., Wang Z. R., Sabbagh S. A., Liu Y. Q., Betti R., Guazzotto L., "Application of benchmarked kinetic resistive wall mode stability codes to ITER, including additional physics", Physics of Plasmas **24** (2017). DOI: 10.1063/1.4989503

[8] Okabayashi M., Bogatu I. N., Chance M. S., Chu M. S., Garofalo A. M., In Y., Jackson G. L., La Haye R. J., Lanctot M. J., Manickam J., Marrelli L., Martin P., Navratil G. A., Reimerdes H., Strait E. J., Takahashi H., Welander A. S., Bolzonella T., Budny R. V., Kim J. S., Hatcher R., Liu Y. Q., Luce T. C., "Comprehensive control of resistive wall modes in DIII-D advanced tokamak plasmas", Nuclear Fusion **49** (2009). DOI: 10.1088/0029-5515/49/12/125003

[9] Zheng L. J., Kotschenreuther M. T., Valanju P., "Investigation of the n = 1 resistive wall modes in the ITER high-mode confinement", Nuclear Fusion **57** (2017). DOI: 10.1088/1741-4326/aa69cb

[10] ITER Research Plan within the Staged Approach (Level III – Provisional Version) ITR-18-003, 2018. ITER Organization. Online: https://www.iter.org/doc/www/content/com/Lists/ITER%20Technical%20Reports/Attachments/9/ITER-Research-Plan_final_ITR_FINAL-Cover_High-Res.pdf Accessed: 2.3.2020

[11] Zanca P., Marrelli L., Manduchi G., Marchiori G., "Beyond the intelligent shell concept: the clean-mode-control", Nuclear Fusion **47** (2007). DOI: 10.1088/0029-5515/47/11/004

[12] Soppelsa A., Marchiori G., Marrelli L., Zanca P., "Design of a new controller of MHD modes in RFX-mod", Fusion Engineering and Design **83** (2008). DOI: 10.1016/j.fusengdes.2008.01.003

[13] Martin P., "Lessons from the RFP on Magnetic Feedback Control of Plasma Stability", Fusion Science and Technology **59** (2011). DOI: 10.13182/FST11-A11700

[14] Zanca P., Marrelli L., Paccagnella R., Soppelsa A., Baruzzo M., Bolzonella T., Marchiori G., Martin P., Piovesan P., "Feedback control model of the m = 2, n = 1 resistive wall mode in a circular plasma", Plasma Physics and Controlled Fusion **54** (2012). DOI: 10.1088/0741-3335/54/9/094004

[15] Hanson J. M., Bialek J. M., Baruzzo M., Bolzonella T., Hyatt A. W., Jackson G. L., King J., La Haye R. J., Lanctot M. J., Marrelli L., Martin P., Navratil G. A., Okabayashi M., Olofsson K. E. J., Paz-Soldan C., Piovesan P., Piron C., Piron L., Shiraki D., Strait E. J., Terranova D., Turco F., Turnbull A. D., Zanca P., "Feedback-assisted extension of the tokamak operating space to low safety factor", Physics of Plasmas **21** (2014). DOI: 10.1063/1.4886796

[16] Brunsell P.R., Yadikin D., Gregoratto D., Paccagnella R., Bolzonella T., Cavinato M., Cecconello M., Drake J.R., Luchetta A., Manduchi G., Marchiori G., Marrelli L., Martin P., Masiello A., Milani F., Ortolani S., Spizzo G., Zanca P., "Feedback Stabilization of Multiple Resistive Wall Modes", Physical Review Letters 93 (2004)

[17] Olofsson K. E. J., Brunsell P. R., Drake J. R., Frassinetti L., "A first attempt at few coils and low-coverage resistive wall mode stabilization of EXTRAP T2R", Plasma Physics and Controlled Fusion **54** (2012). DOI: 10.1088/0741-3335/54/9/094005

[18] Setiadi A. C., Brunsell P. R., Frassinetti L., "Implementation of model predictive control for resistive wall mode stabilization on EXTRAP T2R", Plasma Physics and Controlled Fusion **57** (2015). DOI: 10.1088/0741-3335/57/10/104005

[19] Setiadi A. C., Brunsell P. R., Frassinetti L., "Improved model predictive control of resistive wall modes by error field estimator in EXTRAP T2R", Plasma Physics and Controlled Fusion **58** (2016). DOI: 10.1088/0741-3335/58/12/124002

[20] Setiadi A. C., Brunsell P. R., Villone F., Mastrostefano S., Frassinetti L., "Gray-box modeling of resistive wall modes with vacuum-plasma separation and optimal control design for EXTRAP T2R", Fusion Engineering and Design **121** (2017). DOI: 10.1016/j.fusengdes.2017.07.011

[21] Hanson J. M., De Bono B., Levesque J. P., Mauel M. E., Maurer D. A., Navratil G. A., Sunn Pedersen T., Shiraki D., James R. W., "A Kalman filter for feedback control of rotating external kink instabilities in the presence of noise", Physics of Plasmas **16** (2009). DOI: 10.1063/1.3110110

[22] Strait E. J., Bialek J., Bogatu N., Chance M., Chu M. S., Edgell D., Garofalo A. M., Jackson G. L., Jayakumar R. J., Jensen T. H., Katsuro-Hopkins O., Kim J. S., La Haye R. J., Lao L. L., Makowski M. A., Navratil G., Okabayashi M., Reimerdes H., Scoville J. T., Turnbull A. D., the DIII–D Team, "Resistive wall mode stabilization with internal feedback coils in DIII-D", Physics of Plasmas **11** (2004). DOI: 10.1063/1.1666238

[23] In Y., Kim J. S., Edgell D. H., Strait E. J., Humphreys D. A., Walker M. L., Jackson G. L., Chu M. S., Johnson R., La Haye R. J., Okabayashi M., Garofalo A. M., Reimerdes H., "Model-based dynamic resistive wall mode identification and feedback control in the DIII-D tokamak", Physics of Plasmas **13** (2006). DOI: 10.1063/1.2214637

[24] In Y., Bogatu I. N., Garofalo A.M., Jackson G. L., Kim J. S., La Haye R. J., Lanctot M. J., Marrelli L., Martin P., Okabayashi M., Reimerdes H., Schaffer M. J., Strait E. J., "On the roles of direct feedback and error field correction in stabilizing resistive-wall modes", Nucl. Fusion **50** (2010). DOI: 10.1088/0029-5515/50/4/042001

[25] Hanson J. M., Berkery J. W., Bialek J., Clement M., Ferron J.R., Garofalo A. M., Holcomb C.T., La Haye R. J., Lanctot M. J., Luce T. C., Navratil G. A., Olofsson K. E. J., Strait E. J., Turco F., Turnbull A. D., "Stability of DIII-D high-performance, negative central shear discharges", Nuclear Fusion **57** (2017). DOI: 10.1088/1741-4326/aa6266

[26] Clement M., Hanson J., Bialek J., Navratil G., "GPU-based optimal control for RWM feedback in tokamaks", Control Engineering Practice **68** (2017). DOI: 10.1016/j.conengprac.2017.08.002

[27] Clement M., Hanson J., Bialek J., Navratil G., "H2 optimal control techniques for resistive wall mode feedback in tokamaks", Nuclear Fusion **58** (2018). DOI: 10.1088/1741-4326/aaaecd

[28] Sabbagh S. A., Bell R. E., Menard J. E., Gates D. A., Sontag A. C., Bialek J. M., LeBlanc B. P., Levinton F. M., Tritz K., Yuh H., "Active Stabilization of the Resistive-Wall Mode in High-Beta, Low-Rotation Plasmas", Physical Review Letters **97** (2006). DOI: 10.1103/PhysRevLett.97.045004

[29] Sabbagh S. A., Berkery J. W., Bell R. E., Bialek J. M., Gerhardt S. P., Menard J. E., Betti R., Gates D. A., Hu B., Katsuro-Hopkins O. N., LeBlanc B. P., Levinton F. M., Manickam J., Tritz K., Yuh H., "Advances in global MHD mode stabilization research on NSTX", Nuclear Fusion **50** (2010). DOI: 10.1088/0029-5515/50/2/025020

[30] Sabbagh et al., "Overview of physics results from the conclusive operation of the National Spherical Torus Experiment", Nuclear Fusion **53** (2013). DOI: 10.1088/0029-5515/53/10/104007

[31] Katsuro-Hopkins O., Bialek J., Maurer D. A., Navratil G. A., "Enhanced ITER resistive wall mode feedback performance using optimal control techniques", Nuclear Fusion **47** (2007). DOI:10.1088/0029-5515/47/9/012

[32] Portone A., Villone F., Liu Y., Albanese R., Rubinacci G., "Linearly perturbed MHD equilibria and 3D eddy current coupling via the control surface method". Plasma Physics and Controlled Fusion **50** (2008). DOI: 10.1088/0741-3335/50/8/085004

[33] Albanese R., Liu Y. Q., Portone A., Rubinacci G., Villone F., "Coupling Between a 3-D Integral Eddy Current Formulation and a Linearized MHD Model for the Analysis of Resistive Wall Modes", IEEE Transactions on Magnetics **44** (2008). DOI: 10.1109/TMAG.2007.915303

[34] Villone F., Liu Y., Rubinacci G., Ventre S., "Effects of thick blanket modules on the resistive wall modes stability in ITER", Nuclear Fusion **50** (2010). DOI: 10.1088/0029-5515/50/12/125011

[35] Villone F., Chiariello A. G., Mastrostefano S., Pironti A., Ventre S., "GPU-accelerated analysis of vertical instabilities in ITER including three-dimensional volumetric conducting structures", Plasma Physics and Controlled Fusion **54** (2012). DOI: 10.1088/0741-3335/54/8/085003

[36] Pustovitov V. D., "Decoupling in the problem of tokamak plasma response to asymmetric magnetic perturbations", Plasma Physics and Controlled Fusion **50** (2008). DOI: 10.1088/0741-3335/50/10/105001

[37] Villone F., Barbato L., Mastrostefano S., Ventre S., "Coupling of nonlinear axisymmetric plasma evolution with three-dimensional volumetric conductors", Plasma Physics and Controlled Fusion **55** (2013). DOI: 10.1088/0741-3335/55/9/095008

[38] Barbato L., Formisano A., Martone R., Villone F., "Error Field Impact on Plasma Boundary in ITER Scenarios", IEEE Transactions on Magnetics **52** (2016). DOI: 10.1109/TMAG.2015.2480417

[39] Humphreys D., Ambrosino G., de Vries P., Felici F., Kim S. H., Jackson G., Kallenbach A., Kolemen E., Lister J., Moreau D., Pironti A., Raupp G., Sauter O., Schuster E., Snipes J., Treutterer W., Walker M., Welander A., Winter A., Zabeo L., "Novel aspects of plasma control in ITER", Physics of Plasmas **22** (2015). DOI: 10.1063/1.4907901

[40] Snipes J. A., Albanese R., Ambrosino G., Ambrosino R., Amoskov V., Blanken T. C., Bremond S., Cinque M., de Tommasi G., de Vries P. C., Eidietis N., Felici F., Felton R., Ferron J., Formisano A., Gribov Y., Hosokawa M., Hyatt A., Humphreys D., Jackson G., Kavin A., Khayrutdinov R., Kim D., Kim S. H., Konovalov S., Lamzin E., Lehnen M., Lukash V., Lomas P., Mattei M., Mineev A., Moreau P., Neu G., Nouailletas R., Pautasso G., Pironti A., Rapson C., Raupp G., Ravensbergen T., Rimini F., Schneider M., Travere J.-M., Treutterer W., Villone F., Walker M., Welander A., Winter A., Zabeo L., "Overview of the preliminary design of the ITER plasma control system", Nuclear Fusion **57** (2017). DOI: 10.1088/1741-4326/aa8177

[41] Aydemir A. Y., Park B. H., In Y. K., "Dynamics of an n = 1 explosive instability and its role in high-β disruptions", Nuclear Fusion **58** (2018). DOI: 10.1088/1741-4326/aa95c9

[42] Park J.-K., Boozer A. H., Menard J. E., Schaffer M. J., "Error field correction in ITER", Nuclear Fusion **48** (2008). DOI: 10.1088/0029-5515/48/4/045006

[43] Foussat A., Libeyre P., Mitchell N., Gribov Y., Jong C. T. J., Bessette D., Gallix R., Bauer P., Sahu A., "Overview of the ITER Correction Coils Design", IEEE Transactions on Applied Superconductivity **20** (2010). DOI: 10.1109/TASC.2010.2041911

[44] Foussat A., Weiyue W., Jing W., Shuangsong D., Sgobba S., Hongwei L., Libeyre P., Jong C., Klofac K., Mitchell N., "Mechanical design and construction qualification program on ITERcorrection coils structures", Nuclear Engineering and Design **269** (2014). DOI: 10.1016/j.nucengdes.2013.08.016

[45] Amoskov V. M., Belyakov V. A., Gribov Y. A., Lamzin E. A., Maximenkova N. A., Sytchevsky S. E., "Optimization of Currents in ITER Correction Coils", Physics of Particles and Nuclei Letters **12** (2015). DOI: 10.1134/S1547477115030048

[46] Ariola M., De Tommasi G., Pironti A., Villone F., "Control of resistive wall modes in tokamak plasmas", Control Engineering Practice **24** (2014). DOI: 10.1016/j.conengprac.2013.11.009

[47] Ariola M., Pironti A., "Control of the Resistive Wall Modes for the ITER Tokamak". In: Ariola M., Pironti A., (eds) "Magnetic Control of Tokamak Plasmas", Springer International Publishing Switzerland (2016). DOI: 10.1007/978-3-319-29890-0_11

[48] Loarte A., Huijsmans G., Futatani S., Baylor L. R., Evans T. E., Orlov D. M., Schmitz O., Becoulet M., Cahyna P., Gribov Y., Kavin A., Sashala Naik A., Campbell D. J., Casper T., Daly E., Frerichs H., Kischner A., Laengner R., Lisgo S., Pitts R. A., Saibene G., Wingen A., "Progress on the application of ELM control schemes to ITER scenarios from the non-active phase to DT operation", Nuclear Fusion **54** (2014). DOI: 10.1088/0029-5515/54/3/033007

[49] Choi W., La Haye R. J., Lanctot M. J., Olofsson K. E. J., Strait E. J., Sweeney R., Volpe F. A. and The DIII-D Team, "Feedforward and feedback control of locked mode phase and rotation in DIII-D with application to modulated ECCD experiments", Nuclear Fusion **58** (2018). DOI: 10.1088/1741-4326/aaa6e3

[50] Sen A. K., Nagashima M., Longman R. W., "Optimal control of tokamak resistive wall modes in the presence of noise", Physics of Plasmas **10** (2003). DOI: 10.1063/1.1616560

[51] Sun Z., Sen A. K., Longman R. W., "Adaptive optimal stochastic state feedback control of resistive wall modes in tokamaks", Physics of Plasmas **13** (2006). DOI: 10.1063/1.2161168

[52] Dalessio J., Schuster E., Humphreys D. A., Walker M. L., In Y., Kim J.-S., "Model-based control of the resistive wall mode in DIII-D: A comparison study", Fusion Engineering and Design **84** (2009). DOI: 10.1016/j.fusengdes.2009.01.026

[53] Dalessio J., Schuster E., Humphreys D., Walker M., In Y., Kim J.-S., "Model-Based Robust Control of Resistive Wall Modes via μ Synthesis", Fusion Science and Technology **55** (2009). DOI: 10.13182/FST09-A4069

[54] Villone F., Pironti A., "Effects of power supply limits on control of MHD instabilities in fusion devices", 15th IEEE International Conference on Environment and Electrical Engineering, Rome (2015). DOI: 10.1109/EEEIC.2015.7165212

[55] Maciejowski J., *Predictive Control: With Constraints*, Pearson Education – Prentice Hall (2002). ISBN: 0201398230

[56] Qin S. J., Badgwell T. A., "A survey of industrial model predictive control technology", Control Engineering Practice **11** (2003). DOI: 10.1016/S0967-0661(02)00186-7

[57] Bemporad A., Morari M., Dua V., Pistikopoulos E. N., "The explicit linear quadratic regulator for constrained systems", Automatica **38** (2002). DOI: 10.1016/S0005-1098(01)00174-1

[58] Boyd S., Vandenberghe L., *Convex Optimization*, Cambridge University Press, New York, NY (2004). DOI: 10.1017/CBO9780511804441.

[59] Ferreau H. J., Kirches C., Potschka A., Bock H. G., Diehl M., "qpOASES, a parametric active-set algorithm for quadratic programming", Mathematical Programming Computation **6** (2014). DOI: 10.1007/s12532-014-0071-1

[60] Mattingley J. Boyd S., "CVXGEN: a code generator for embedded convex optimization", Optimization and Engineering **13** (2012). DOI: 10.1007/s11081-011-9176-9

[61] Hartley E. N., Jerez J. L., Suardi A., Maciejowski J. M., Kerrigan E. C., Constantinides G. A., "Predictive control using an FPGA

[61] with application to aircraft control", IEEE Transactions on Control Systems Technology **22** (2014). DOI: 10.1109/TCST.2013.2271791

[62] Richter S., Jones C. N., Morari M., "Computational Complexity Certification for Real-Time MPC With Input Constraints Based on the Fast Gradient Method", IEEE Transactions on Automatic Control **57** (2012). DOI: 10.1109/TAC.2011.2176389

[63] Giselsson P., "Improved fast dual gradient methods for embedded model predictive control", IFAC Proceedings Volumes **47** (2014). DOI: 10.3182/20140824-6-ZA-1003.00295

[64] Jerez J. L., Goulart P. J., Richter S., Constantinides G. A., Kerrigan E. C., Morari M. "Embedded online optimization for model predictive control at megahertz rates", IEEE Transactions on Automatic Control **59** (2014). DOI 10.1109/TAC.2014.2351991

[65] Maljaars E., Felici F., de Baar M. R., van Dongen J., Hogeweij G. M. D., Geelen P. J. M., Steinbuch M., "Control of the tokamak safety factor profile with time-varying constraints using MPC", Nuclear Fusion **55** (2015). DOI: 10.1088/0029-5515/55/2/023001

[66] Maljaars E., Felici F., Blanken T. C., Galperti C., Sauter O., de Baar M. R., Carpanese F., Goodman T. P., Kim D., Kim S. H., Kong M., Mavkov B., Merle A., Moret J. M., Nouailletas R., Scheffer M., Teplukhina A. A., Vu N. M. T., The EUROfusion MST1-team, The TCV-team, "Profile control simulations and experiments on TCV: a controller test environment and results using a model-based predictive controller", Nuclear Fusion **57** (2017). DOI: 10.1088/1741-4326/aa8c48

[67] Wehner W., Lauret M., Schuster E., Ferron J. R., Holcomb C., Luce T. C., Humphreys D. A., Walker M. L., Penaflor B. G., Johnson R. D., "Predictive control of the tokamak q profile to facilitate reproducibility of high-$q_{min}$ steady-state scenarios at DIII-D", 2016 IEEE Conference on Control Applications (CCA), Buenos Aires, Argentina (2016). DOI: 10.1109/CCA.2016.7587900

[68] Gerkšič S., Pregelj B., Perne M., Knap M., De Tommasi G., Ariola M., Pironti A., "Plasma current and shape control for ITER using fast online MPC", 20th IEEE Real Time Conference, Padova (2016). DOI: 10.1109/RTC.2016.7543092

[69] Gerkšič S., Pregelj B., Perne M., Ariola M., De Tommasi G., Pironti A., "Model predictive control of ITER plasma current and shape using singular-value decomposition", Fusion Engineering and Design **129** (2018). DOI: 10.1016/j.fusengdes.2018.01.074

[70] Gerkšič S., De Tommasi G., "Vertical stabilization of ITER plasma using explicit model predictive control", Fusion Engineering and Design **88** (2013). DOI: 10.1016/j.fusengdes.2013.02.021

[71] Song I., Liu H., Li J., Gao G., Daly E., Tao J., "Conceptual Design of ITER In-Vessel Vertical Stabilization Coil Power Supply System", IEEE Transactions on Applied Superconductivity **24** (2014). DOI: 10.1109/TASC.2013.2292671

[72] Nesterov Y., "Gradient methods for minimizing composite functions", Mathematical Programming **140** (2013). DOI: 10.1007/s10107-012-0629-5

[73] Davison E., "A new method for simplifying large linear dynamic systems", IEEE Transactions on Automatic Control **13** (1968). DOI: 10.1109/TAC.1968.1098866

[74] Anderson B. D. O., Moore J. B., "Optimal control: linear quadratic methods". Prentice Hall, Englewood Cliffs (1990). ISBN: 0486457664

[75] Gerkšič S., Strmčnik S., van den Boom, T., "Feedback action in predictive control: An experimental case study". Control Engineering Practice **16** (2008). DOI: 10.1016/j.conengprac.2007.04.012

[76] Kvasnica, M.: "Real-time model predictive control via multi-parametric programming". VDM verlag, Saarbrücken, 2009. ISBN: 3639206444

[77] Su W., Boyd S., Candès E. J., "A differential equation for modeling Nesterov's accelerated gradient method: theory and insights", The Journal of Machine Learning Research **15**, 2016. URL: http://jmlr.org/papers/v17/15-084.html